\documentclass[onecolumn,amsmath,showpacs,11pt]{revtex4}
\usepackage{graphicx}
\usepackage{dcolumn}
\usepackage{bm}
\begin{document}
\newcommand{\hs}{\hspace*{0.5cm}}
\newcommand{\vs}{\vspace*{0.5cm}}
\newcommand{\be}{\begin{equation}}
\newcommand{\ee}{\end{equation}}
\newcommand{\bea}{\begin{eqnarray}}
\newcommand{\eea}{\end{eqnarray}}
\newcommand{\ben}{\begin{enumerate}}
\newcommand{\een}{\end{enumerate}}
\newcommand{\bde}{\begin{widetext}}
\newcommand{\ede}{\end{widetext}}
\newcommand{\nn}{\nonumber}
\newcommand{\crn}{\nonumber \\}
\newcommand{\Tr}{\mathrm{Tr}}
\newcommand{\non}{\nonumber}
\newcommand{\noi}{\noindent}
\newcommand{\al}{\alpha}
\newcommand{\la}{\lambda}
\newcommand{\bet}{\beta}
\newcommand{\ga}{\gamma}
\newcommand{\va}{\varphi}
\newcommand{\om}{\omega}
\newcommand{\pa}{\partial}
\newcommand{\+}{\dagger}
\newcommand{\fr}{\frac}
\newcommand{\bc}{\begin{center}}
\newcommand{\ec}{\end{center}}
\newcommand{\Ga}{\Gamma}
\newcommand{\de}{\delta}
\newcommand{\De}{\Delta}
\newcommand{\ep}{\epsilon}
\newcommand{\varep}{\varepsilon}
\newcommand{\ka}{\kappa}
\newcommand{\La}{\Lambda}
\newcommand{\si}{\sigma}
\newcommand{\Si}{\Sigma}
\newcommand{\ta}{\tau}
\newcommand{\up}{\upsilon}
\newcommand{\Up}{\Upsilon}
\newcommand{\ze}{\zeta}
\newcommand{\ps}{\psi}
\newcommand{\Ps}{\Psi}
\newcommand{\ph}{\phi}
\newcommand{\vph}{\varphi}
\newcommand{\Ph}{\Phi}
\newcommand{\Om}{\Omega}

\title{Kinetic mixing effect in the 3-3-1-1 model}

\author{P. V. Dong}
\email {pvdong@iop.vast.ac.vn} \affiliation{Institute of Physics, Vietnam Academy of Science and Technology, 10 Dao Tan, Ba Dinh, Hanoi, Vietnam}
\author{D. T. Si}
\email {dtsi@grad.iop.vast.ac.vn} \affiliation{Institute of Physics, Vietnam Academy of Science and Technology, 10 Dao Tan, Ba Dinh, Hanoi, Vietnam}
\date{\today}

\begin{abstract}
We show that the mixing effect of the neutral gauge bosons in the 3-3-1-1 model comes from two sources. The first one is due to the 3-3-1-1 gauge symmetry breaking as usual, whereas the second one results from the kinetic mixing between the gauge bosons of $U(1)_X$ and $U(1)_N$ groups, which are used to determine the electric charge and baryon minus lepton numbers, respectively. Such mixings modify the $\rho$-parameter and the known couplings of $Z$ with fermions. The constraints that arise from flavor-changing neutral currents due to the gauge boson mixings and non-universal fermion generations are also given.  
\end{abstract}


\maketitle

\section{Introduction}

The standard model is incomplete since it leaves crucial questions of the nature unsolved, namely the neutrino masses, dark matter, matter-antimatter asymmetry, cosmic inflation, and so on \cite{pdg}. Many such difficulties of the standard model can be solved by the recently-proposed $SU(3)_C\otimes SU(3)_L\otimes U(1)_X\otimes U(1)_N$ (3-3-1-1) gauge model, where $SU(3)_C$ is the ordinary color group, $SU(3)_L$ is an extension of the weak-isospin symmetry ($SU(2)_L$), and the last two factors correspondingly define the electric charge ($Q$) and baryon-minus-lepton charge ($B-L$), respectively~\cite{3311,d3311}. This is the most simple framework that unifies the electroweak and $B-L$ interactions in a nontrivial way, analogously to the electroweak theory. The new model also provides insights in the electric charge quantization (which is due to the $B-L$ dynamics in general \cite{d3311}, while only the minimal 3-3-1-1 versions have additional quantization condition that results from specific fermion contents like the 3-3-1 models \cite{ecq}) and flavor questions (where the dangerous FCNCs due to the unwanted vacuums and interactions are suppressed by $W$-parity conservation \cite{3311}, whereas the contribution of $U(1)_N$ gauge boson including the kinetic mixing effect discussed below could relax those 3-3-1 model's bounds for the B physics anomalies \cite{flav331}). 

The 3-3-1-1 model contains four neutral gauge bosons, the photon, $Z$, and new $Z'$, $Z''$. Their mixing effects due to the 3-3-1-1 gauge symmetry breaking have been studied \cite{3311,d3311}. However, since this theory includes two $U(1)$ factor groups, the kinetic mixing \cite{kineticmixing} between the corresponding gauge bosons is unavoidable, which might cause significant effects and modify the well-measured parameters/observables. It has not been examined yet. In this work, we interpret this mixing and investigate its corrections to the known parameters and constraints. The correlation between the two kinds of mixings is also evaluated.            

\section{\label{model} The 3-3-1-1 model and kinetic mixing}

Assume that all the left-handed fermion doublets of $SU(2)_L$ are enlarged to the fundamental representations of $SU(3)_L$ (i.e., triplets or antitriplets), while all the right-handed fermion singlets of $SU(2)_L$ by themselves transform as singlets of $SU(3)_L$. The $SU(3)_L$ anomaly cancellation requires the number of fermion triplets is equal that of fermion antitriplets. Thus, the fermion content of the 3-3-1-1 model under consideration is given by \cite{d3311} 
\bea && \psi_{aL}\equiv 
\left(\begin{array}{c}
\nu_{aL}\\
e_{aL}\\
k_{aL}\end{array}\right)  \sim \left(1,3,\fr{-1+q}{3},\fr{-2+n}{3}\right),\\
&& \nu_{aR}\sim \left(1,1,0,-1\right),\hs e_{aR}\sim (1,1,-1,-1),\hs k_{aR}\sim (1,1,q,n),\\
&& Q_{3 L}\equiv 
\left(\begin{array}{c}
u_{3L}\\
d_{3L}\\
j_{3L}\end{array}\right)   \sim \left(3,3,\fr{1+q}{3},\fr{2+n}{3}\right),\hs Q_{\al L}\equiv 
\left(\begin{array}{c}
d_{\al L}\\
-u_{\al L}\\
j_{\al L} \end{array}\right)  \sim \left(3,3^*,-\fr{q}{3},-\fr{n}{3}\right),\\
&& u_{a R}\sim \left(3,1,\fr 2 3,\fr 1 3\right),\hs d_{aR}\sim \left(3,1,-\fr 1 3,\fr 1 3\right),\\
&& j_{3R}\sim \left(3,1,\fr 2 3 +q,\fr 4 3 +n\right),\hs j_{\al R}\sim \left(3,1,-\fr 1 3 -q,-\fr 2 3 -n\right), \eea
where $a=1,2,3$ and $\al=1,2$ are generation indices, the quantum numbers in the parentheses are defined upon the 3-3-1-1 symmetries, respectively, and the new fields $k_{aL,R}$, $j_{aL,R}$, and $\nu_{aR}$ have been included to complete the representations and cancel the other anomalies. For special cases, $k_{aR}$ are excluded while $k_{aL}$ are replaced by either $(e_{aR})^c$ or $(\nu_{aR})^c$, called minimal 3-3-1-1 versions, respectively. But, this does not work for quarks since the symmetries, $SU(3)_C$, $SU(3)_L$, and space-time, commute. Hence, $j_a$ are necessarily introduced. Note also that the following discussions generally apply for all cases. The $Q$ and $B-L$ charges of the new fermions are \bea && Q(\nu_R)=0,\hs Q(k)=q,\hs Q(j_3)=\fr 2 3+q,\hs Q(j_\al)=-\fr 1 3 -q,\\
&& [B-L](\nu_R)=-1,\hs [B-L](k)=n,\hs [B-L](j_3)=\fr 4 3 +n,\hs [B-L](j_\al)=-\fr 2 3 -n. \eea We see that $(q,n)$ are those charges defined for $k_a$ fields, which satisfy $-2.08011 < q < 1.08011$ in order to have a correct, effective Weinberg angle as explicitly shown below, and $n\neq (2m-1)/3$ for any integer $m$ to have a nontrivial, residual, discrete symmetry of the gauge symmetry, which stabilizes the dark matter candidates.     

To break the 3-3-1-1 symmetry and generate appropriate masses for the particles, the scalar content contains \cite{d3311} 
\bea \eta &=&
\left(
\begin{array}{l}
\eta^{0,0}_1\\
\eta^{-1,0}_2\\
\eta^{q,n+1}_3
\end{array}\right)\sim \left(1,3,\fr{q-1}{3},\fr{n+1}{3}\right),\hs
\rho =
\left(
\begin{array}{l}
\rho^{1,0}_1\\
\rho^{0,0}_2\\
\rho^{q+1,n+1}_3
\end{array}\right)\sim \left(1,3,\fr{q+2}{3},\fr{n+1}{3}\right),\\
\chi &=&
\left(
\begin{array}{l}
\chi^{-q,-n-1}_1\\
\chi^{-q-1,-n-1}_2\\
\chi^{0,0}_3
\end{array}\right)\sim\left(1,3,-\fr{2q+1}{3},-\fr 2 3 (n+1)\right),\hs
\phi \sim (1,1,0,2),\eea where the superscripts define $(Q,B-L)$ values respectively, while the subscripts indicate $SU(3)_L$ component fields. The corresponding vacuum expectation values (VEVs) are obtained as  
\bea \langle \eta \rangle &=& \fr{1}{\sqrt{2}}\left(
\begin{array}{c}
u \\
0\\
0
\end{array}\right),\hs
\langle \rho\rangle =
\fr{1}{\sqrt{2}} \left(
\begin{array}{c}
0\\
v \\
0
\end{array}\right),\hs
\langle \chi\rangle =
\fr{1}{\sqrt{2}} \left(
\begin{array}{c}
0\\
0\\
w
\end{array}\right),\hs \langle \phi\rangle = \fr{1}{\sqrt{2}} \La.\eea  
The VEVs $w, u, v$ break the 3-3-1-1 symmetry to $SU(3)_C\otimes U(1)_Q\otimes U(1)_{B-L}$, while the VEV $\La$ breaks $B-L$ to a discrete symmetry, $U(1)_{B-L}\rightarrow P$. The residual operators can be identified as
\bea && Q=T_3+\beta T_8+X,\hs  B-L=\beta' T_8+N,\hs  P=(-1)^{3(\beta' T_8 + N)+2s},\eea
where $\beta = -(1+2q)/\sqrt{3}$, $\beta'=-2(1+n)/\sqrt{3}$, and $s$ is spin. Note that $\beta$ is bounded by $-1.82455<\beta<1.82455$ due to the $q$ constraint, and $\beta'\neq \fr{4m}{3\sqrt{3}}$ for any $m$ integer. The weak hypercharge is $Y=\beta T_8+X$. Furthermore, because $w,\La$ give the masses for the new particles, whereas $u,v$ are for the ordinary particles, to be consistent with the standard model, we assume $u,v\ll w,\La$.   

Up to the gauge fixing and ghost terms, the total Lagrangian takes the form,
\bea \mathcal{L}&=&\sum_{F} \bar{F}i\ga^\mu D_\mu F + \sum_{S}(D^\mu S)^\dagger (D_\mu S) + \mathcal{L}_{\mathrm{Yukawa}}- V(\eta,\rho,\chi,\phi)\crn
&&-\fr 1 4G_{i\mu\nu}G_i^{\mu\nu} -\fr 1 4  A_{i\mu\nu} A_i^{\mu\nu} -\fr 1 4 B_{\mu\nu}B^{\mu\nu} -\fr 1 4 C_{\mu\nu}C^{\mu\nu}-\fr{\delta}{2}  B_{\mu\nu}C^{\mu\nu},\eea  where $F$ and $S$ run over all fermion multiplets and scalar multiplets respectively, and $\delta$ is a dimensionless parameter. $\mathcal{L}_{\mathrm{Yukawa}}$ and $V(\eta,\rho,\chi,\phi)$ are Yukawa Lagrangian and scalar potential respectively, which their explicit forms are easily obtained. The covariant derivative and field strength tensors are given by   
\bea D_\mu &=& \pa_\mu + i g_s t_i G_{i\mu} + i g T_i A_{i\mu} + i g_X X B_\mu+ i g_N N C_\mu,\\
G_{i\mu\nu}&=&\pa_\mu G_{i\nu}-\pa_\nu G_{i\mu}-g_s f_{ijk} G_{j\mu} G_{k\nu},\\
A_{i\mu\nu}&=&\pa_\mu A_{i\nu}-\pa_\nu A_{i\mu}-g f_{ijk} A_{j\mu} A_{k\nu},\\
B_{\mu\nu}&=&\pa_\mu B_\nu-\pa_\nu B_\mu,\hs C_{\mu\nu}=\pa_\mu C_\nu-\pa_\nu C_\mu,\eea where $\{g_s,\ g,\ g_X,\ g_N\}$, $\{t_i,\ T_i,\ X,\ N\}$, and $\{G_i,\ A_i,\ B,\ C\}$ stand for coupling constants, generators, and gauge bosons of the 3-3-1-1 groups, respectively. Notably, the $\delta$ term, called kinetic mixing, was omitted in the previous studies, in spite of the fact that it is gauge invariant and also cannot be transformed away by rescaling the gauge fields. Even if its tree-level value vanishes, it can be radiatively induced. The existence of the $\delta$ term is a new observation of this work. We should impose $|\delta|<1$ in order to have a definitely positive kinetic energy.    

Because of the kinetic mixing term, the two $U(1)$ gauge bosons $B_\mu$ and $C_\mu$ are generally not orthogonal and normalized. Let us rewrite the kinetic terms of $B_\mu$ and $C_\mu$ as
\be \mathcal{L}=\cdots -\fr 1 4 B^2_{\mu\nu}-\fr 1 4 C^2_{\mu\nu}-\fr{\delta}{2}  B_{\mu\nu}C^{\mu\nu}=\cdots -\fr 1 4 (B_{\mu\nu}+\delta C_{\mu\nu})^2-\fr 1 4 (1-\delta^2) C^2_{\mu\nu},\ee which takes the canonical form by a non-unitary transformation $(B_\mu,C_\mu)\rightarrow (B'_\mu,C'_\mu)$ as
\be B'=B+\delta C,\hs C'=\sqrt{1-\delta^2}C.\ee Substituting, $C=\fr{1}{\sqrt{1-\delta^2}}C'$ and $B=B'-\fr{\delta}{\sqrt{1-\delta^2}}C'$, into the covariant derivative, it becomes       
\be 
D_\mu = \pa_\mu + i g_s t_i G_{i\mu} + i g T_i A_{i\mu} + i g_X X B'_\mu+\fr{i}{\sqrt{1-\delta^2}}(g_N N-g_X X\delta)  C'_\mu, 
\ee which is given in terms of the physical (orthogonal and normalized) fields $(B'_\mu,C'_\mu)$. 

The 3-3-1-1 gauge symmetry breaking also leads to mixings among $A_{3}$, $A_8$, $B'$, and $C'$. Their mass Lagrangian arises from $\sum_S (D_\mu\langle S\rangle)^\dagger (D^\mu\langle S\rangle)$, which yields  
\bea 
 \mathcal{L}_{\mathrm{mass}}^{\mathrm{neutral}}  =
  \fr1 2\left(A_3 \  A_8  \  B' \ C'  \right) M^2 
 \left(\begin{array}{c}
   A_3\\
   A_8 \\
   B' \\
   C'  
\end{array} \right), \eea 
 where the mass matrix $M^2$ is symmetric and its elements $M^2=\{m^2_{ij}\}$ are given by 
 \bea
m^2_{11}&= & \fr {g^2} 4 (u^2 + v^2),\hs m^2_{12}=  \fr {g^2} {4\sqrt3} (u^2-v^2),\hs m^2_{13} = -\fr{g^2t_X}{4\sqrt3}\left[(\sqrt3+\beta)u^2+(\sqrt3-\beta)v^2\right], \crn 
m^2_{14}&=& \fr{g^2}{4\sqrt3\sqrt{1-\delta^2}}\left\{[\delta (\sqrt3+\beta)t_X-\beta' t_N] u^2 +[\delta (\sqrt3-\beta)t_X+\beta' t_N] v^2 \right\} ,\crn
m^2_{22}&=&\fr {g^2} {12}(u^2+v^2+4w^2),\hs
m^2_{23}= -\fr{g^2t_X}{12}\left[(\sqrt3+\beta)u^2-(\sqrt3-\beta)v^2+4\beta w^2\right], 
\crn
m^2_{24}&=& \fr{g^2} { 12 \sqrt{1-\delta^2}}\left\{[\delta (\sqrt3+\beta)t_X-\beta' t_N] u^2-[\delta (\sqrt3-\beta)t_X+\beta' t_N] v^2 +4(\delta\beta t_X-\beta' t_N)w^2\right\},\crn
m^2_{33} &=&   \fr{g^2t^2_X}{12}\left[(\sqrt3+\beta)^2u^2 +(\sqrt3-\beta)^2v^2 + 4\beta^2w^2\right],\crn
m^2_{34}&=& \fr{-g^2 t_X } { 12\sqrt{1-\delta^2}}\left\{ (\sqrt3+\beta)[\delta (\sqrt3+\beta)t_X-\beta' t_N] u^2+ (\sqrt3-\beta)[\delta (\sqrt3-\beta)t_X+\beta' t_N] v^2 \right.\crn
&&\left.+4\beta \left(\delta\beta t_X-\beta'  t_N\right)w^2\right\}, \crn
m^2_{44}&=& \fr{g^2} { 12(1-\delta^2)}\left\{[\delta t_X(\sqrt3+\beta) - \beta' t_N]^2 u^2+[\delta t_X(\sqrt3-\beta) + \beta'  t_N]^2  v^2  +4(\delta\beta t_X-\beta'  t_N)^2w^2\right.\crn
&& \left. +48(1-\delta^2)t^2_N\La^2\right\},\nn
\eea where we have defined $t_X=g_X/g$ and $t_N=g_N/g$.

It is easily obtained that $M^2$ has a zero eigenvalue (the photon mass) with corresponding eigenstate (the photon field), given by   
\be
 A = s_W A_3 + c_W \left(\beta t_W A_8 +\sqrt{1-\beta^2 t^2_W}B'\right),\label{exexex}\ee where $s_W=e/g=t_X/\sqrt{1+(1+\beta^2)t_X^2}$ is the sine of the Weinberg angle, which can explicitly be identified from electromagnetic interaction vertices \cite{donglong}. The last relation further implies $s^2_W<1/(1+\beta^2)$ or $|\beta|<\cot_W$, which yields the mentioned $q$ bounds, since $\beta=-(1+2q)/\sqrt{3}$ and $s^2_W=0.231$, effectively given at the weak scales $(u,v)$ \cite{d3311}. Note that the field in the parenthesis of (\ref{exexex}) is properly coupled to the weak hypercharge $Y=\beta T_8 + X$. Therefore, we can define the standard model $Z$ and new $Z'$ as follows 
 \bea 
Z&=&c_W A_3 - s_W \left(\beta t_W A_8 +\sqrt{1-\beta^2 t^2_W}B'\right),\\
Z' &=&\sqrt{1-\beta^2 t^2_W} A_8 - \beta t_W B',\eea which are given orthogonally to $A$, as usual. At this stage, $C'$ is always orthogonal to $A,Z,Z'$.       
 
Let us change to the new basis $(A_3,A_8,B',C')\rightarrow (A,Z,Z',C')$,\bea
\left(\begin{array}{c}
A_3\\
A_8 \\
B' \\
C' \end{array}\right) = U_1 \left( \begin{array}{c}
A\\
Z \\
Z' \\
C' \end{array}\right), \hs 
 U_1 = \left (\begin{array}{cccc}
 s_W &c_W&0& 0\\
\beta s_W &-\beta s_W t_W &\sqrt{1-\beta^2t^2_W}&0\\
c_W\sqrt{1-\beta^2t^2_W}&- s_W\sqrt{1-\beta^2 t^2_W}&- \beta t_W&0\\
0&0&0&1
\end{array} \right ). \label{neutral2}
\eea 
The mass matrix $M^2$ becomes then 
\bea M'^2 = U^T_1 M^2 U_1=
\left(
\begin{array}{cc}
0 & 0 \\
0 & M'^2_{s}\end{array} \right),\hs M'^2_s \equiv
\left(
\begin{array}{ccc}
m^2_{Z} & m^2_{ZZ'} & m^2_{ZC'}\\
m^2_{ZZ'} & m^2_{Z'} & m^2_{Z'C'}\\
m^2_{ZC'} & m^2_{Z'C'} & m^2_{C'}
\end{array}\right). 
\eea We see that the photon field is physical and decoupled, while $Z,Z',C'$ mix via the $3\times 3$ mass submatrix $M'^2_s$ with the elements given by 
 \bea
 m^2_Z&=&\fr {g^2}{4c^2_W} (u^2 + v^2),\hs m^2_{ZZ'} =\fr{g^2\left[(1+\sqrt3\beta t^2_W)u^2-(1-\sqrt3\beta t^2_W)v^2\right]}{4\sqrt3 c_W\sqrt{1-\beta^2 t^2_W}},\nn\eea \bea 
m^2_{ZC'}&=& \fr{g^2}{4\sqrt3 c_W \sqrt{1-\delta^2}}\left\{\left[\fr{\delta t_W(\sqrt3+\beta)}{\sqrt{1-\beta^2 t^2_W}}-\beta' t_N\right]u^2+\left[\fr{\delta t_W (\sqrt3 - \beta)}{\sqrt{1-\beta^2 t^2_W}} + \beta' t_N\right]v^2\right\},\nn\eea \bea 
 m^2_{Z'}&= &\fr{g^2\left[(1+\sqrt3\beta t^2_W)^2u^2+(1-\sqrt3\beta t^2_W)^2v^2 +4 w^2 \right]}{12(1-\beta^2 t^2_W)},\nn\eea \bea 
 m^2_{Z'C'}&= &\fr{g^2}{12\sqrt{(1-\delta^2)(1-\beta^2 t^2_W)}}\left\{(1+\sqrt3\beta t^2_W)\left[\fr{\delta t_W(\sqrt3+\beta)}{\sqrt{1-\beta^2 t^2_W}}-\beta' t_N\right] u^2\right.\crn 
 &&-\left.(1-\sqrt3\beta t^2_W)\left[\fr{\delta t_W(\sqrt3-\beta)}{\sqrt{1-\beta^2 t^2_W}}+\beta' t_N\right] v^2+4\left[\fr{\delta\beta t_W}{\sqrt{1-\beta^2 t^2_W}}-\beta' t_N\right]w^2\right\},\nn \eea \bea
  m^2_{C'}&= & \fr{g^2}{12(1-\delta^2)}\left\{\left[ \fr{\delta t_W(\sqrt3+\beta)}{\sqrt{1-\beta^2 t^2_W}}-\beta' t_N\right]^2 u^2+\left[ \fr{\delta t_W(\sqrt3-\beta)}{\sqrt{1-\beta^2 t^2_W}}+\beta' t_N\right]^2v^2 \right. \crn
  &&\left.+4\left[\fr{\delta\beta t_W}{\sqrt{1-\beta^2 t^2_W}}-\beta' t_N \right]^2w^2+48(1-\delta^2)t^2_N\La^2 \right\}. \nn
 \eea

Because of the condition $u,v\ll w,\La$, we have $m^2_{Z}, m^2_{ZZ'}, m^2_{ZC'}\ll m^2_{Z'}, m^2_{Z'C'},m^2_{C'}$. The mass matrix $M'^2_s$ can be diagonalized by using the seesaw formula \cite{seesaw} to separate the light state $Z$ from the heavy states $Z',C'$, which is given by   
 \bea
\left(\begin{array}{c}
A\\
Z \\
Z' \\
C' \end{array}\right) = U_2 \left(\begin{array}{c}
A\\
Z_1 \\
\mathcal{Z}' \\
\mathcal{C}'\end{array}\right), \hs M''^2=U^{T}_2M'^2U_2= \left(\begin{array}{ccc}
0 &0& 0 \\
0 &m^{2}_{Z_{1}} & 0\\
0 & 0&M''^2_{s}\end{array}\right),
\eea
where $Z_1$ is physical and decoupled, while $\mathcal{Z}'$ and $\mathcal{C}'$ mix via $M''^2_s$, and  
\bea
&& U_2 \simeq  \left(\begin{array}{ccc}
1 & 0 & 0  \\
0 &  1 & \mathcal{E} \\
0 & -\mathcal{E}^T & 1 \\
\end{array}
\right), \hs \mathcal{E} \equiv (m^2_{ZZ'}\ m^2_{ZC'})\left(\begin{array}{cc}
m^2_{Z'} & m^2_{Z'C'}\\
m^2_{Z'C'} & m^2_{C'}
\end{array}
\right)^{-1}, \\
&& m^2_{Z_1}\simeq  m^2_{Z}- \mathcal{E} \left(\begin{array}{cc}
m^2_{ZZ'}\\
m^2_{ZC'} \end{array}\right),\hs 
M''^2_{s} \simeq    \left(\begin{array}{cc}
m^2_{Z'}&m^2_{Z'C'}\\
m^2_{Z'C'}&m^2_{C'} \end{array}\right). 
\eea
Further, we approximate $\mathcal{E}_i\simeq \mathcal{E}^0_i+\mathcal{E}^{\delta}_i$ due to $u,v\ll w,\La$ again, where $i=1,2$, and   
 \bea 
\mathcal{E}^0_1 & = & \dfrac{\sqrt{1-\beta^2t^2_W}}{16c_W\La^2\om^2}  \left\{ 4\sqrt3\left[(1+\sqrt3\beta t^2_W)u^2 -(1-\sqrt3\beta t^2_W)v^2\right] \La^2 + \beta \beta'^2 t^2_W(u^2+v^2)\om^2\right\},\\
 \mathcal{E}^0_2 & = & \dfrac{\beta \beta' t^2_W (u^2+v^2)}{16c_Wt_N\La^2},\hs \mathcal{E}^{\delta}_1  =  \fr{\delta^2}{1-\delta^2}\dfrac{ \beta \beta'^2t^2_W\sqrt{1-\beta^2 t^2_W}(u^2+v^2)}{16c_W\La^2},\\
 \mathcal{E}^{\delta}_2 & = & \fr{\delta}{\sqrt{1-\delta^2}}\dfrac{t_W\sqrt{1-\beta^2 t^2_W}(u^2+v^2)}{16c_Wt^2_N \La^2} +\fr{\delta^2}{1+\sqrt{1-\delta^2}-\delta^2}\dfrac{\beta \beta' t^2_W (u^2+v^2)}{16c_Wt_N\La^2}. 
 \eea
The parameters $\mathcal{E}^0_i$ determine the mixings of $Z$ with $Z',C'$ due to the gauge symmetry breaking, which are like those in the ordinary 3-3-1-1 model without the kinetic mixing \cite{3311,d3311}. Whereas, the parameters $\mathcal{E}^{\delta}_i$ characterize those mixings coming from the kinetic mixing term. Because $\delta$ is finite, the magnitudes of the two $\mathcal{E}^0_i$ and $\mathcal{E}^{\delta}_i$ contributions are equivalent. However, all of them, $\mathcal{E}^{0}_i$ and $\mathcal{E}^{\delta}_i$, are negligibly small due to $u,v\ll w,\La$.        

Finally, it is easily to diagonalize the mass matrix $M''^2_s$ to yield two remaining physical gauge bosons, called $Z_2$ and $Z_3$, such that  
 \bea
 && \left(\begin{array}{c}
A\\
Z_1 \\
\mathcal{Z}' \\
\mathcal{C}' \end{array}\right) = U_3 \left(\begin{array}{c}
A\\
Z_1 \\
Z_2 \\
Z_3\end{array}\right), \hs U_3=\left(\begin{array}{cccc}
1 & 0 & 0 & 0\\
0 & 1 & 0 & 0\\
0 & 0 & c_\xi & s_\xi \\
0 & 0 & -s_\xi & c_\xi 
\end{array}\right),\crn
&& M'''^2=U^T_3 M''^2 U_3=\mathrm{diag}(0,m^2_{Z_1},m^2_{Z_2},m^2_{Z_3}).
 \eea The $\mathcal{Z}'$-$\mathcal{C}'$ mixing angle and $Z_{2}$, $Z_3$ masses are given by  
 \bea 
&& t_{2\xi}  \simeq \dfrac{2\sqrt{1-\delta^2}\left(\delta\beta t_W-\beta' \sqrt{1-\beta^2 t^2_W}t_N\right)w^2}{12(1-\delta^2)(1-\beta^2 t^2_W)\La^2 t^2_N+\left[(\delta\beta t_W-\beta' t_N\sqrt{1-\beta^2 t^2_W})^2- (1-\delta^2)\right]w^2},\\
&& m^2_ {Z_2,Z_3}= \fr1 2 \left[m^2_ {Z'} +m^2_{C'} \mp \sqrt{(m^2_ {Z'} - m^2_{C'})^2 + 4m^4_ {Z' C'} }\right]. 
 \eea It is clear that the kinetic mixing term also contributes to the $\mathcal{Z}'$ and $\mathcal{C}'$ mixing angle with a magnitude equivalent to that due to the gauge symmetry breaking. Unlike the previous case, these mixings are radically large, supposed that $\La\sim w$. However, the mixing effects on the new neutral gauge bosons will cancel when $\xi=0$, or 
 \be \delta =\fr{\beta' t_N}{\beta t_X},\ee where note that $t_X=t_W/\sqrt{1-\beta^2 t^2_W}$.   
 
At this stage, let us summarize that the canonical gauge states are related to the mass eigenstates, $(A_3\ A_8\ B'\ C')^T=U' (A\ Z_1\ Z_2\ Z_3)^T$, by the unitary matrix $U'=U_1U_2U_3$, while the original gauge states are connected to the mass eigenstates, $(A_3\ A_8\ B\ C)^T=U (A\ Z_1\ Z_2\ Z_3)^T$, by the non-unitary matrix $U=U_\delta U'$, where 
 \bea U_{\delta}=\left(\begin{array}{cccc}
1 & 0 & 0 & 0\\
0 & 1 & 0 & 0\\
0 & 0 & 1 & -\fr{\delta}{\sqrt{1-\delta^2}} \\
0 & 0 & 0 & \fr{1}{\sqrt{1-\delta^2}} 
\end{array}\right).\eea The fields $A$, $Z_1$ can be identified like those of the standard model, whereas $Z_2$ and $Z_3$ are the new, heavy neutral gauge bosons, originating from $Z'$ of the 3-3-1 model \cite{331m,331r} and $C$ of $U(1)_N$ symmetry. The mixings of the standard model gauge bosons with the new gauge particles are small due to $|\mathcal{E}_{1,2}|\ll1$ or $u,v\ll w,\La$, while the mixings within the new gauge bosons $Z'$ and $C$ may be large since $\xi$ is finite, provided that $w\sim\La$.            

\section{$\rho$ and $\mathcal{E}_{1,2}$ parameters}

The $\rho$-parameter (or $\Delta\rho\equiv \rho-1$ used below) that is due to the contribution of the new physics comes from two distinct sources, denoted as $\Delta \rho = (\Delta \rho)_{\mathrm{tree}}+(\Delta \rho)_{\mathrm{rad}}$, where the first term results from the tree-level mixing of $Z$ with $Z'$ and $C'$, while the second term originates from the dominant, radiative correction of a heavy non-Hermitian gauge doublet $(X^{-q,-1-n}=\fr{A_4-iA_5}{\sqrt{2}},\ Y^{-1-q,-1-n}=\fr{A_6-iA_7}{\sqrt{2}})$, similarly to the 3-3-1 model case \cite{sturho}. The two kinds of these contributions are suppressed by $(u^2,v^2)/(w^2,\La^2)$, which can become comparable, and are given by 
  \bea 
(\Delta\rho)_{\mathrm{tree}} &=& \fr{m^2_W}{c^2_Wm^2_{Z_1}}-1=\fr{m^2_Z}{m^2_Z-\mathcal{E}(m^2_{ZZ'}\ m^2_{ZC'})^T}-1\simeq \mathcal{E}(m^2_{ZZ'}\ m^2_{ZC'})^T/m^2_Z\crn
 &\simeq & \fr{\left[(1+\sqrt3\beta t^2_W)u^2-(1-\sqrt3\beta t^2_W)v^2\right]^2}{4 (u^2+v^2) w^2}+\fr{\beta^2\beta'^{2}t^4_W (u^2+v^2)} {16(1-\delta^2)\La^2} + \fr{ \delta t_W}{16\sqrt3(1-\delta^2)c_Wt^2_N}\crn
 &&\times \left\{\sqrt3\left(1+\fr{\beta\beta' t_Wt_N}{\sqrt{1-\beta^2 t^2_W}}\right) \fr{u^2+v^2}{\La^2}+\left(\beta-\beta' t_N +\fr{\beta^2\beta' t_Wt_N}{\sqrt{1-\beta^2 t^2_W}}\right)\fr{u^2-v^2}{\La^2}\right\},\\  
 (\Delta \rho)_{\mathrm{rad}}&=&\fr{3\sqrt{2}G_F}{16\pi^2}\left(m^2_Y+m^2_X-\fr{2m^2_Y m^2_X}{m^2_Y-m^2_X}\ln\fr{m^2_Y}{m^2_X}\right)\crn
 &&+\fr{\al}{4\pi s^2_W}\left(\fr{m^2_Y+m^2_X}{m^2_Y-m^2_X}\ln\fr{m^2_Y}{m^2_X}-2-\sqrt{3}\beta t^2_W\ln\fr{m^2_Y}{m^2_X}\right),\eea where $m^2_W=\fr{g^2}{4}(u^2+v^2)$, $m^2_X=\fr{g^2}{4}(u^2+w^2)$, $m^2_Y=\fr{g^2}{4}(v^2+w^2)$, $\sqrt{2}G_F=1/(u^2+v^2)$, and $\al=g^2s^2_W/(4\pi)$. Note that the mass of $W$ boson implies $u^2+v^2=v^2_{\mathrm{w}}=(246\ \mathrm{GeV})^2$. 
 
We first remark that if $\La\gg w$, $\Delta\rho$ depends only on $\beta$ and $w$, not on $\La$, $t_N$, $\beta'$, and $\delta$, which is analogous to that of the corresponding 3-3-1 model. If $\La \sim w$, all the mentioned parameters contribute to $\Delta\rho$. In this case, without loss of generality, we will take $\La=2w$ and $t_N=0.5$ into account. To set other inputs, we are primarily interested in the 3-3-1-1 models that provide dark matter candidates, so $q=0$ or $q=-1$, i.e. $\beta=-1/\sqrt{3}$ or $\beta=1/\sqrt{3}$, respectively \cite{d3311}. Also, the candidates are stabilized if $P$ is nontrivial, as mentioned, so let $n=0$, thus $\beta'=-2/\sqrt{3}$, for simplicity. Furthermore, we would also be concerned with the 3-3-1-1 models that have a low Landau pole \cite{landau}, such that $q=1$ or $q=-2$, thus $\beta=-\sqrt{3}$ or $\beta=\sqrt{3}$, respectively. Although, these two models possess the distinct new physics regimes, we will only  investigate the one with $q=1$, so $\beta=-\sqrt{3}$. From the global fit, the $\rho$-parameter is bounded by $0.00016<\Delta \rho<0.00064$~\cite{pdg}. Since $u^2+v^2=(246\ \mathrm{GeV})^2$, $u$ will vary from 0 to 246 GeV, while $v$ is followed. 

\begin{figure}[!h]
\begin{center}
\includegraphics[scale=0.4]{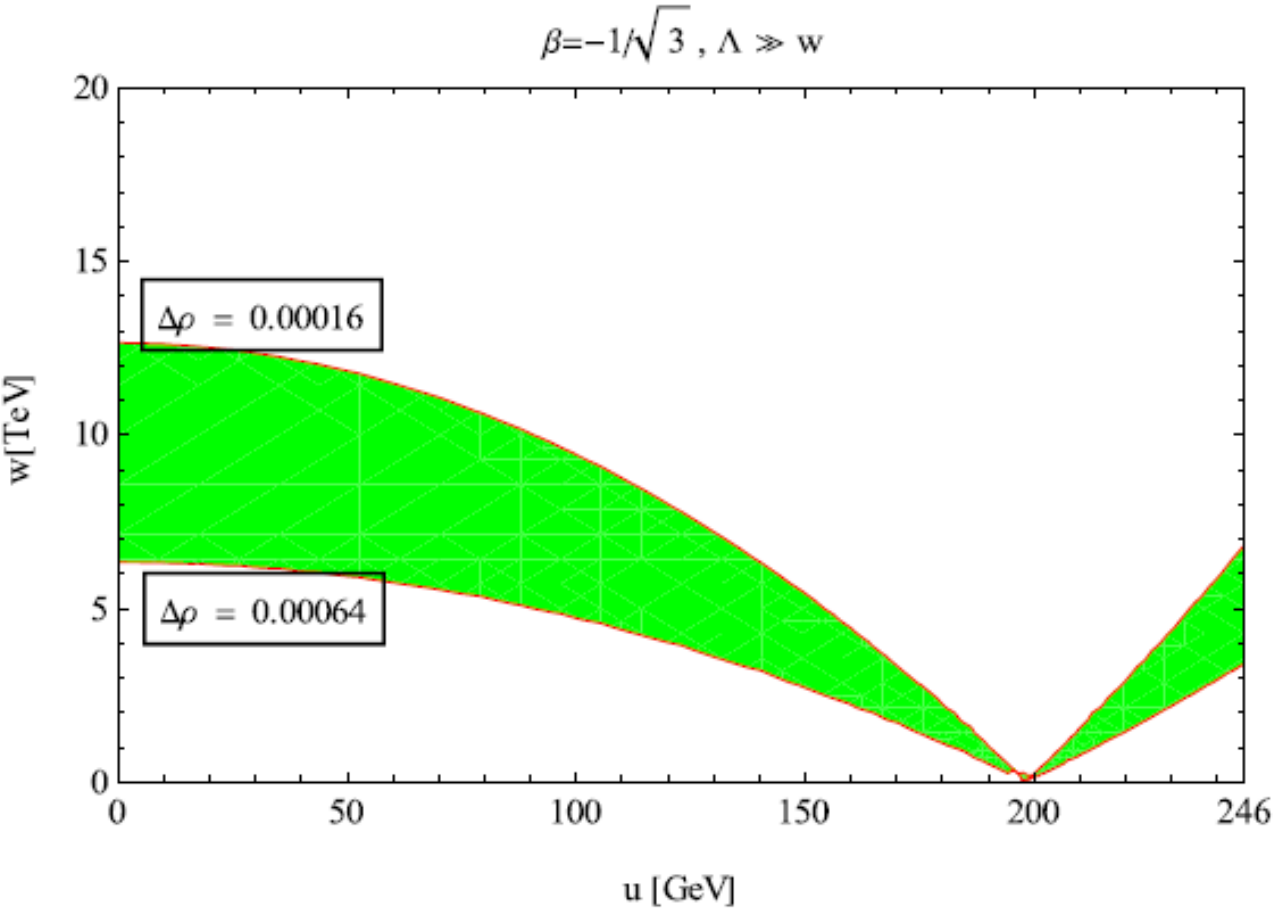}
\includegraphics[scale=0.4]{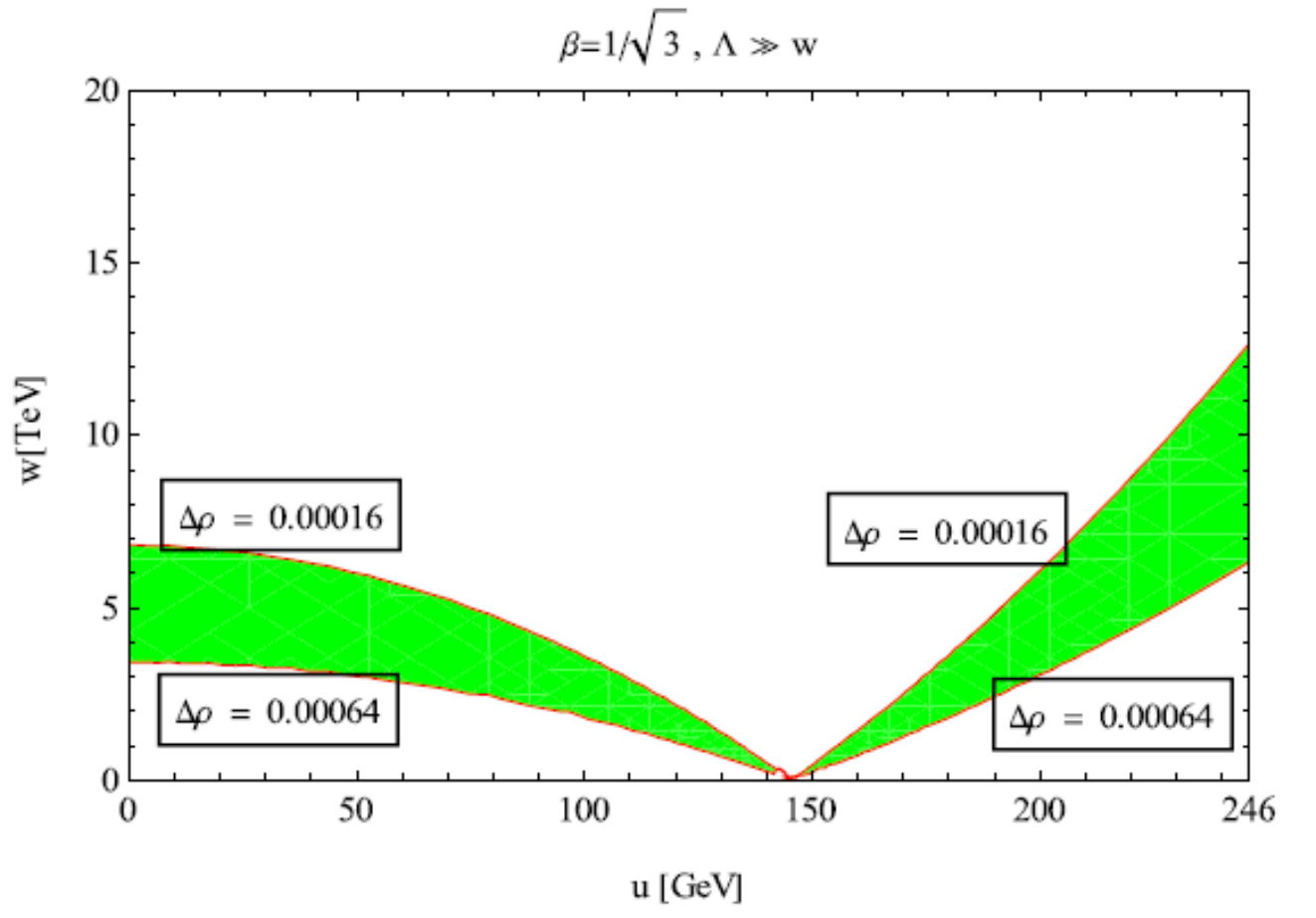}
\includegraphics[scale=0.45]{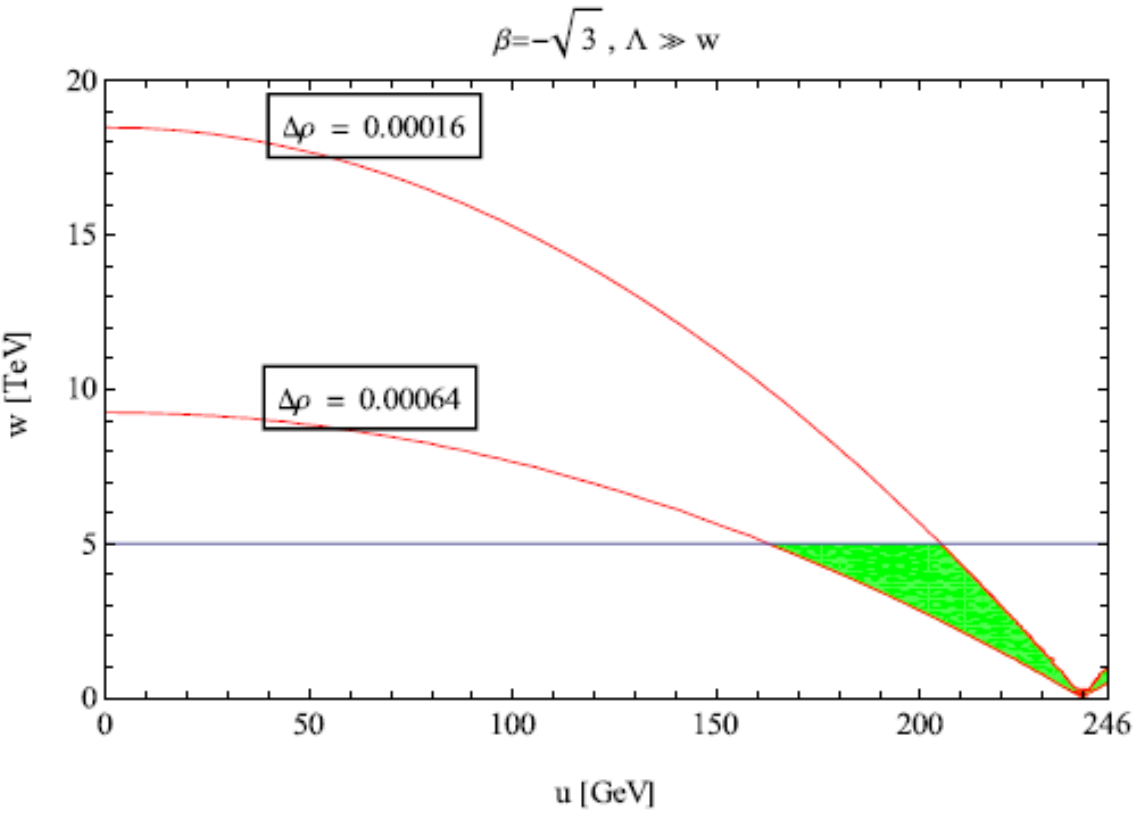}
\caption[]{\label{rho331} The $(u,w)$ regime that is bounded by the $\rho$ parameter for $\beta=-1/\sqrt{3}$, $\La\gg w$ (left panel), $\beta=1/\sqrt{3}$, $\La\gg w$ (middle panel), and $\beta=-\sqrt{3}$, $\La\gg w$ (right panel). For the last case, the Landau pole, e.g. $w=5$ TeV, should be imposed.}
\end{center}
\end{figure}

\begin{figure}[!h]
\begin{center}
\includegraphics[scale=0.4]{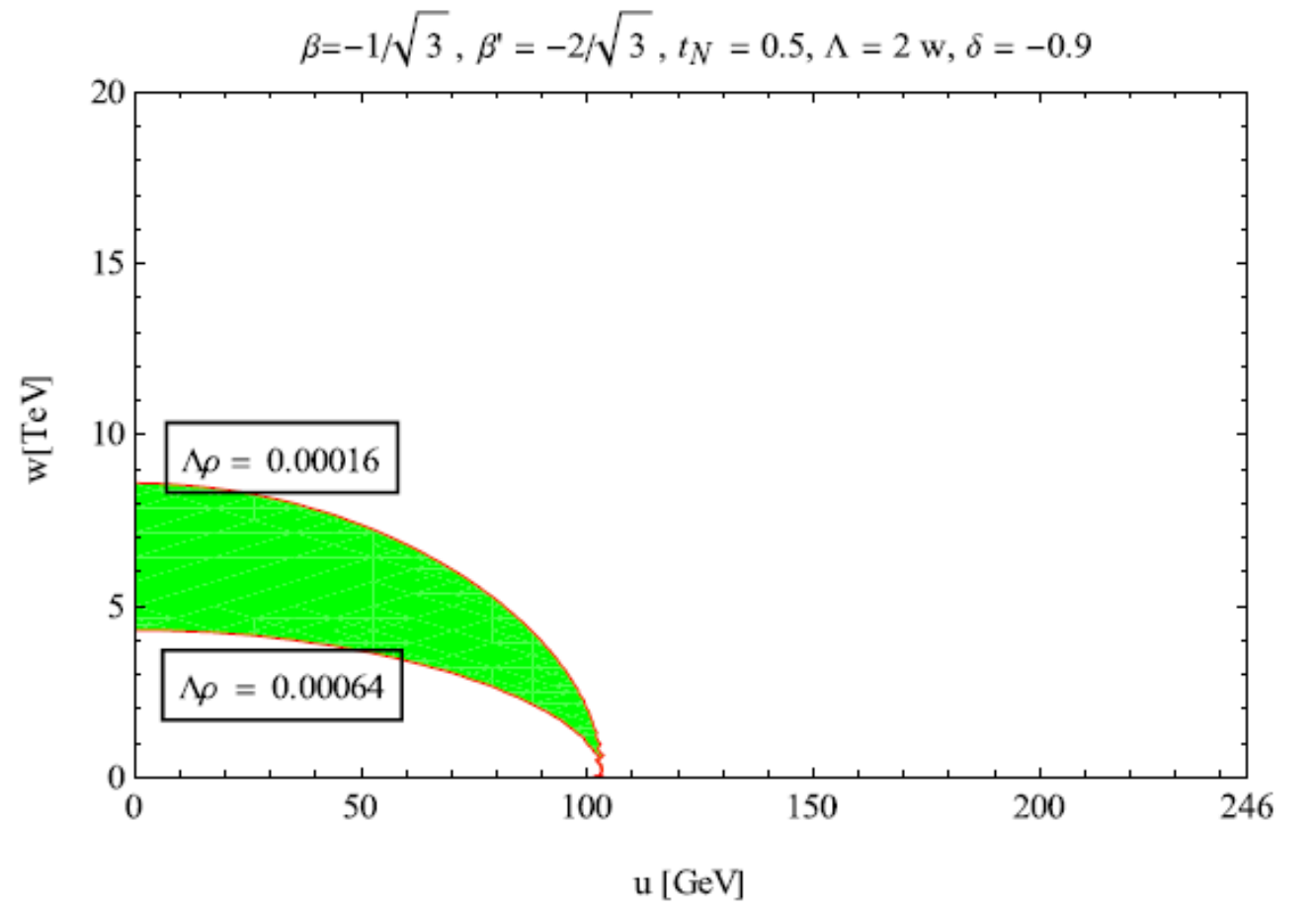}
\includegraphics[scale=0.4]{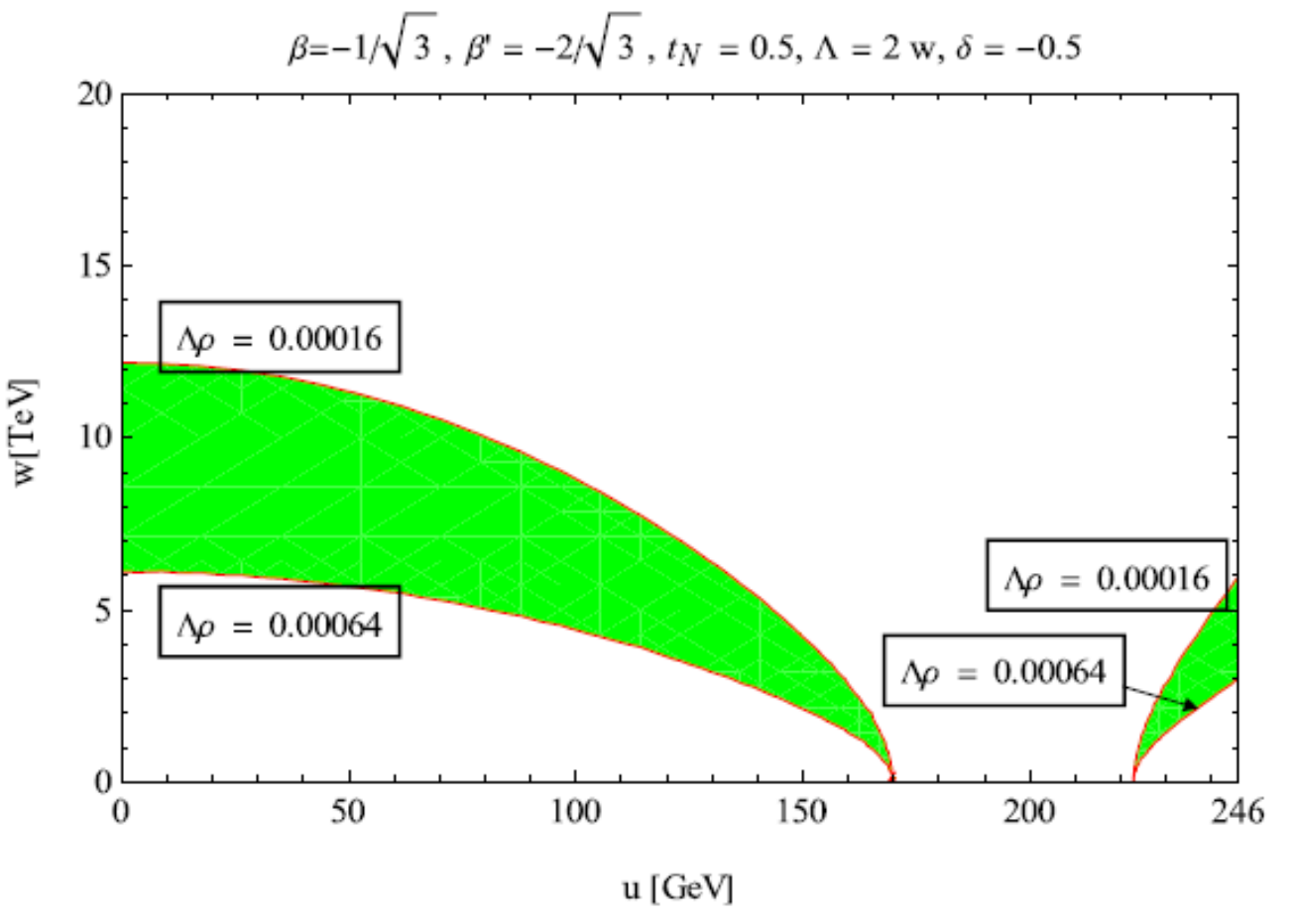}
\includegraphics[scale=0.4]{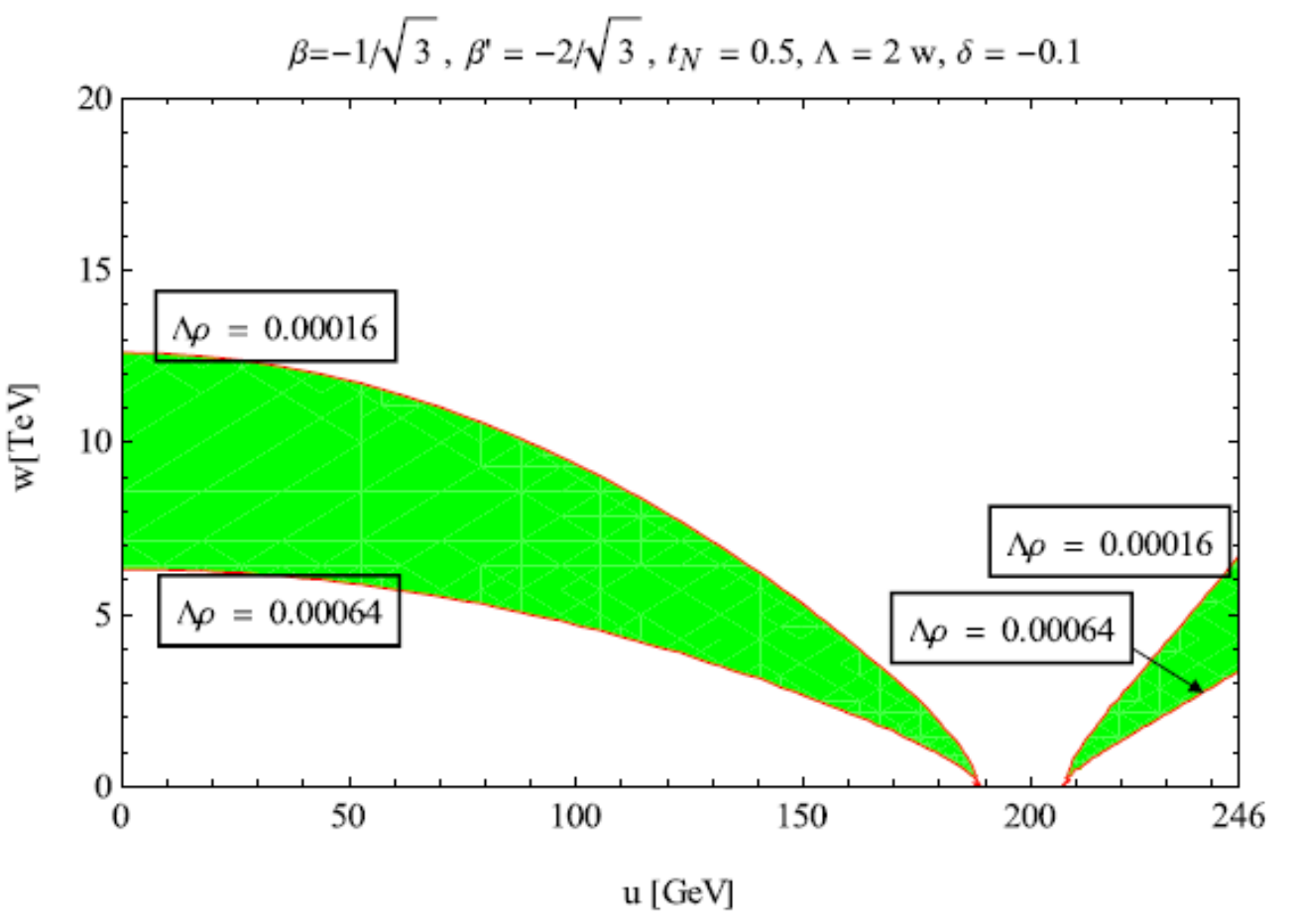}
\includegraphics[scale=0.4]{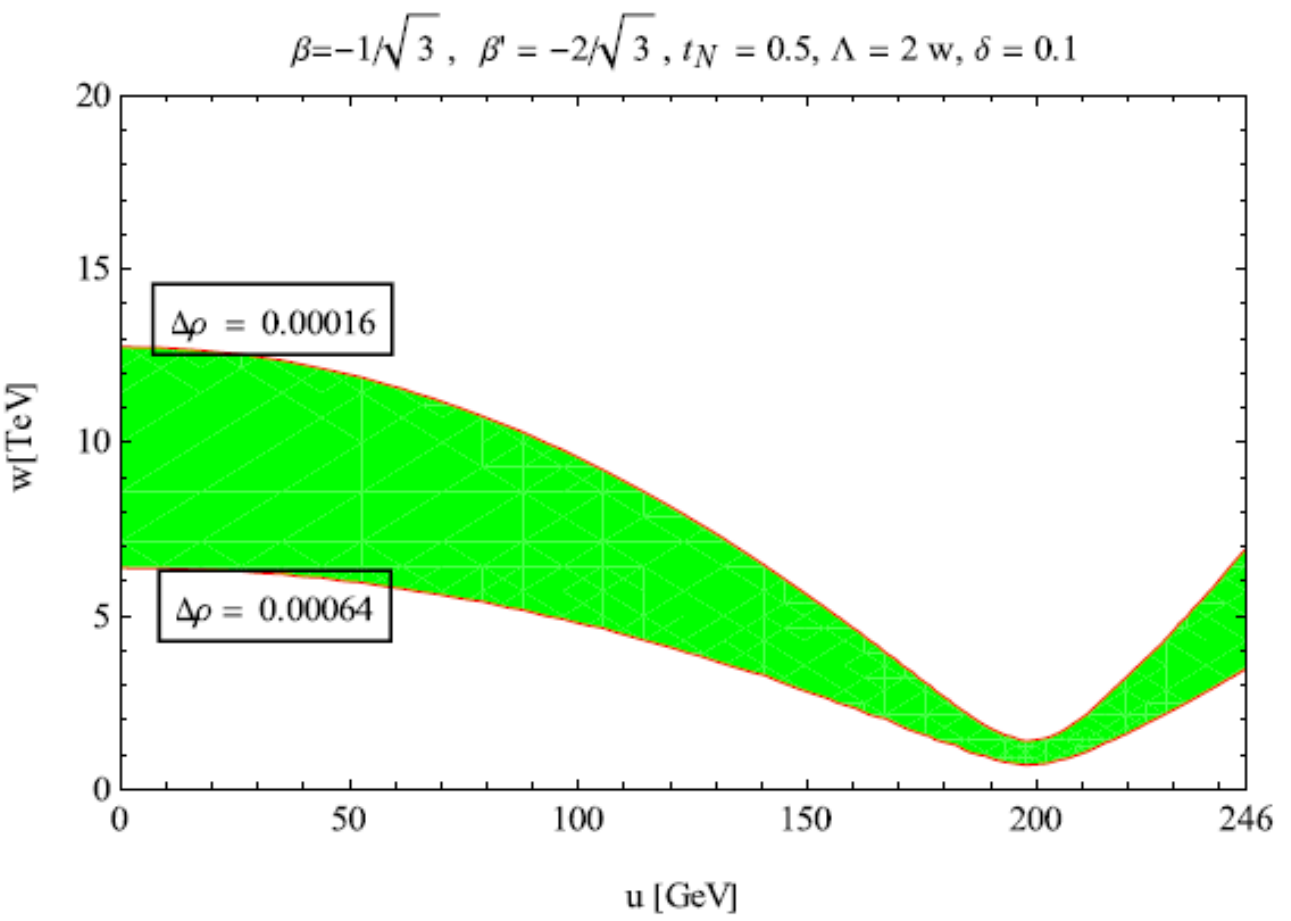}
\includegraphics[scale=0.4]{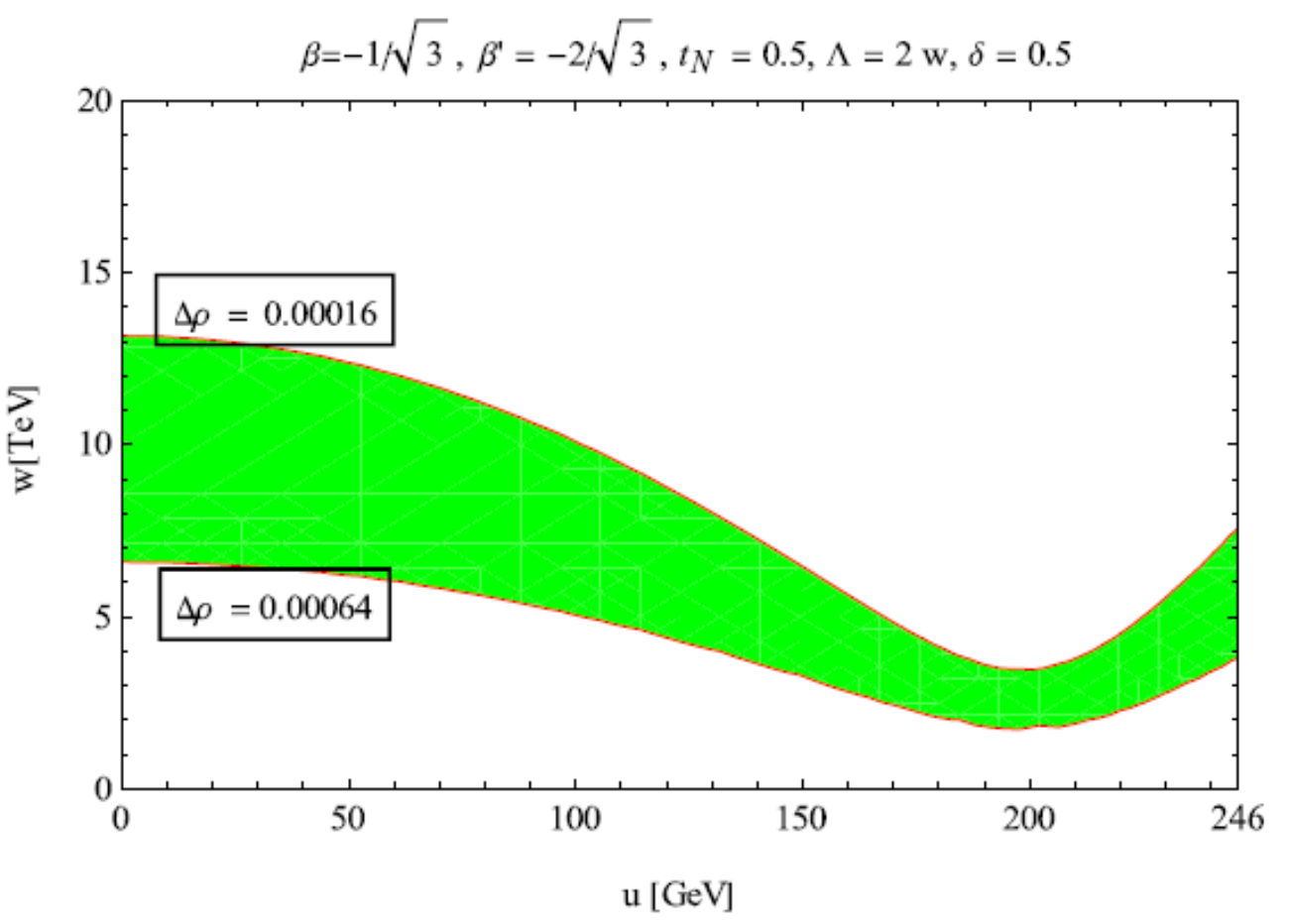}
\includegraphics[scale=0.45]{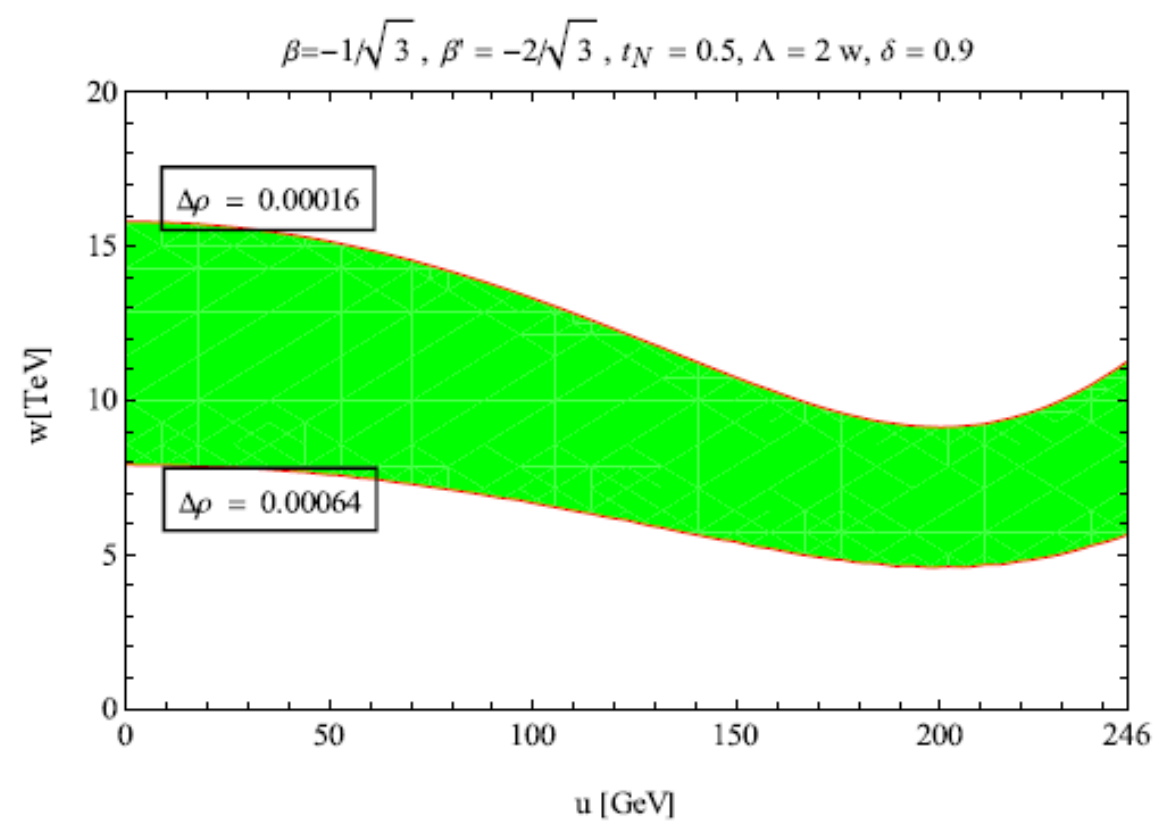}
\caption[]{\label{rho3311r} The $(u,w)$ regime that is bounded by the $\rho$ parameter for $\beta=-1/\sqrt{3}$, $\beta'=-2/\sqrt{3}$, $t_N=0.5$, and $\La=2w$, where the panels, ordering from left to right in raws and then from the top raw down to the bottom raw, correspond to $\delta=-0.9,\ -0.5,\ -0.1,\ 0.1,\ 0.5$, and $0.9$, respectively.}
\end{center}
\end{figure}

\begin{figure}[!h]
\begin{center}
\includegraphics[scale=0.45]{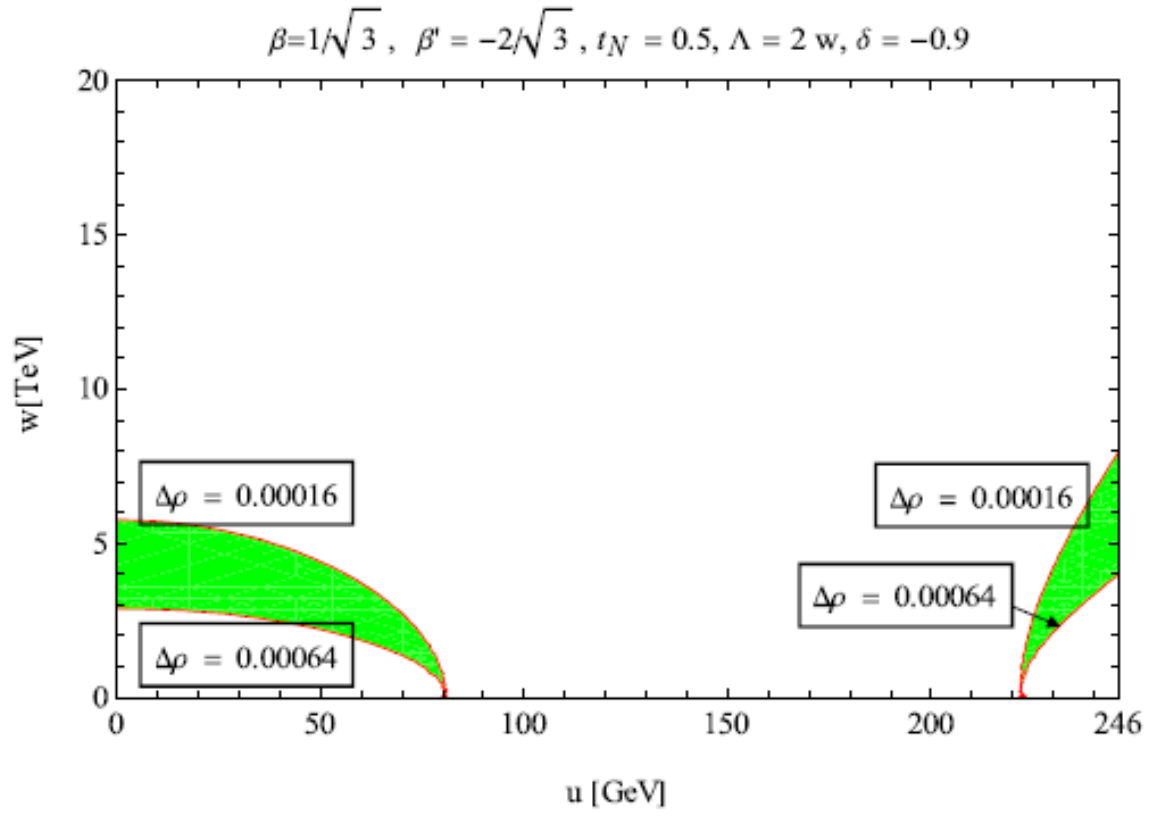}
\includegraphics[scale=0.4]{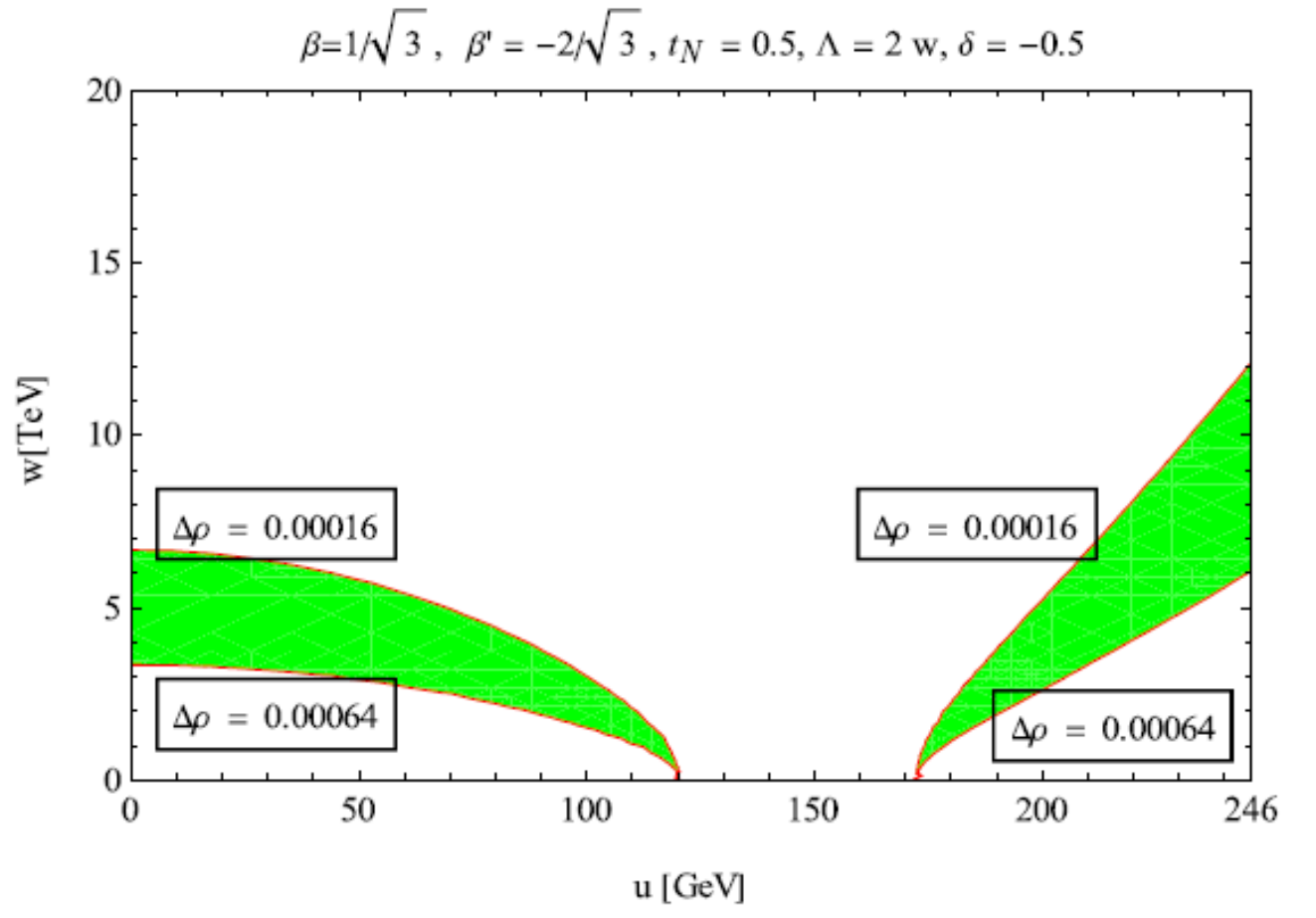}
\includegraphics[scale=0.4]{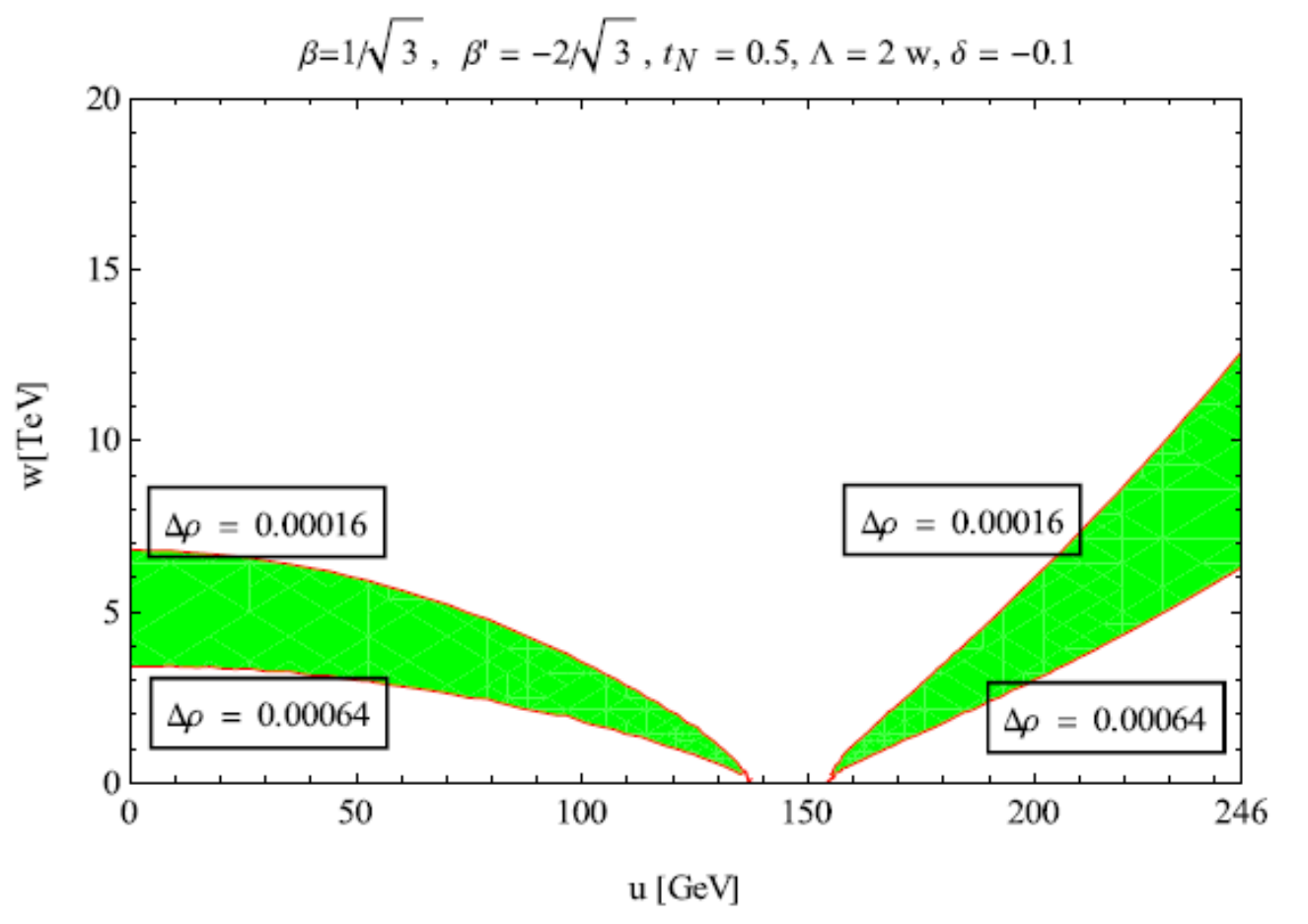}
\includegraphics[scale=0.45]{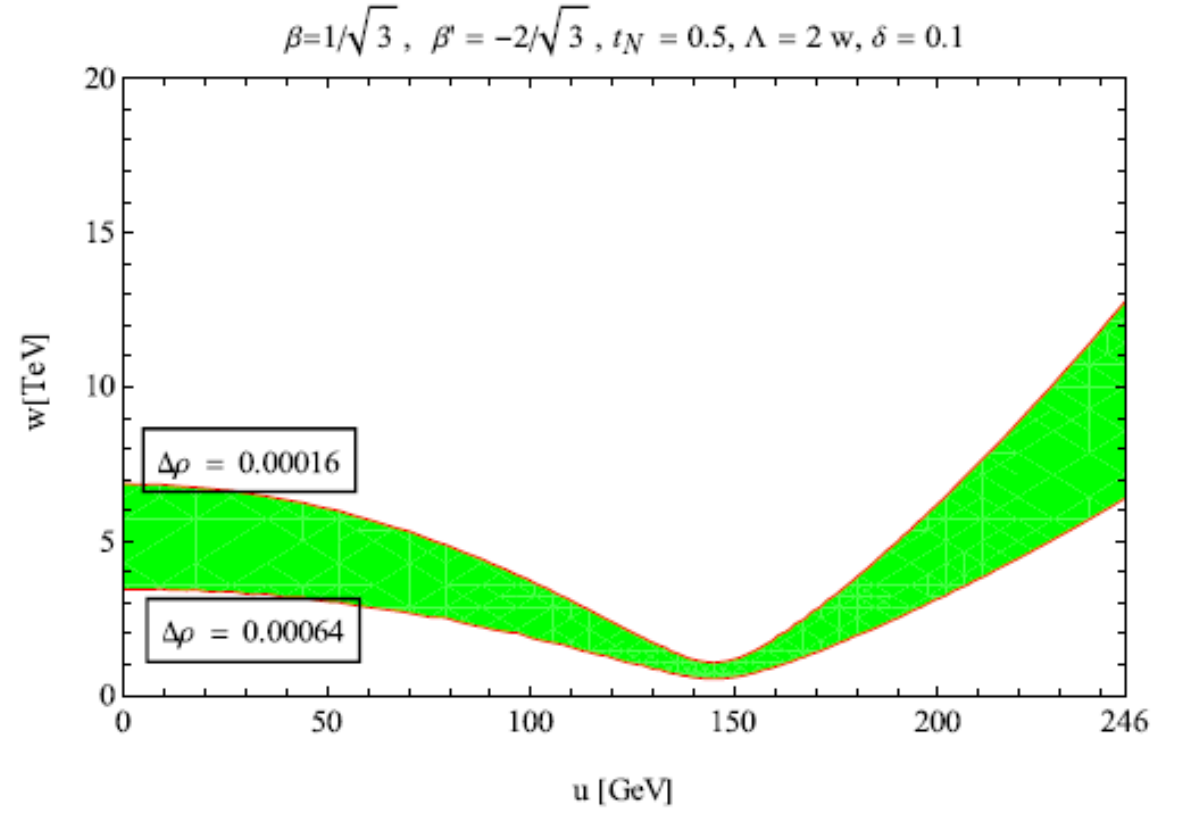}
\includegraphics[scale=0.4]{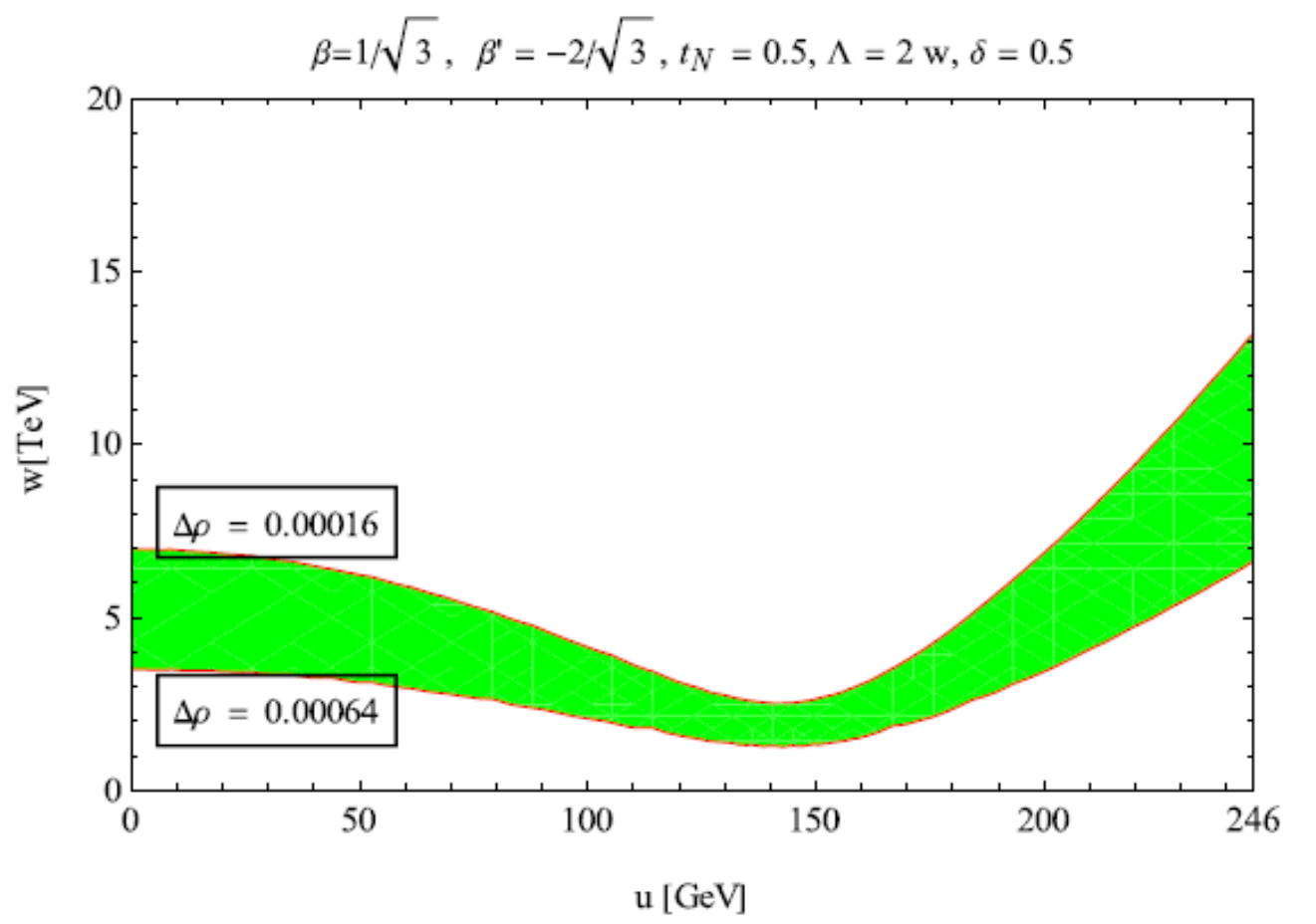}
\includegraphics[scale=0.45]{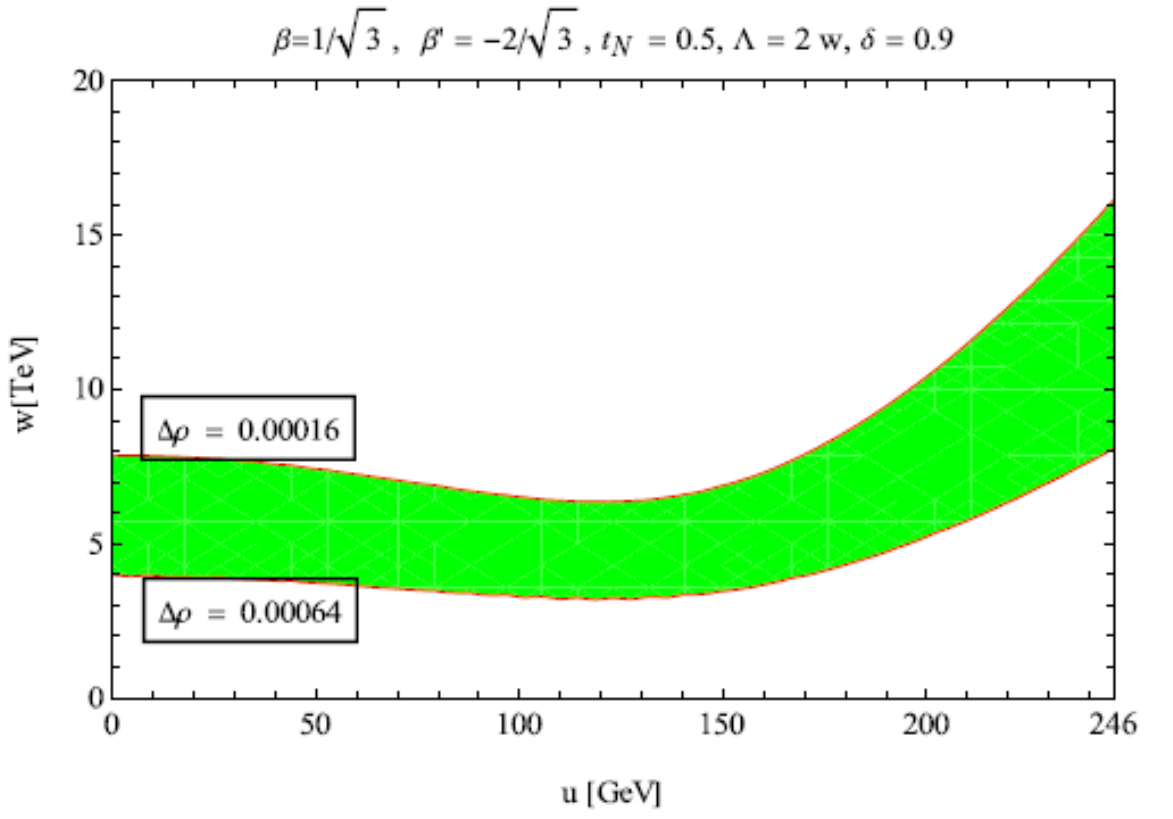}
\caption[]{\label{rho3311s} The $(u,w)$ regime that is bounded by the $\rho$ parameter for $\beta=1/\sqrt{3}$, $\beta'=-2/\sqrt{3}$, $t_N=0.5$, and $\La=2w$, in which the panels, reading from left to right in raws and from the top raw down to the bottom raw, are for $\delta=-0.9,\ -0.5,\ -0.1,\ 0.1,\ 0.5$, and $0.9$, respectively.}
\end{center}
\end{figure}

\begin{figure}[!h]
\begin{center}
\includegraphics[scale=0.4]{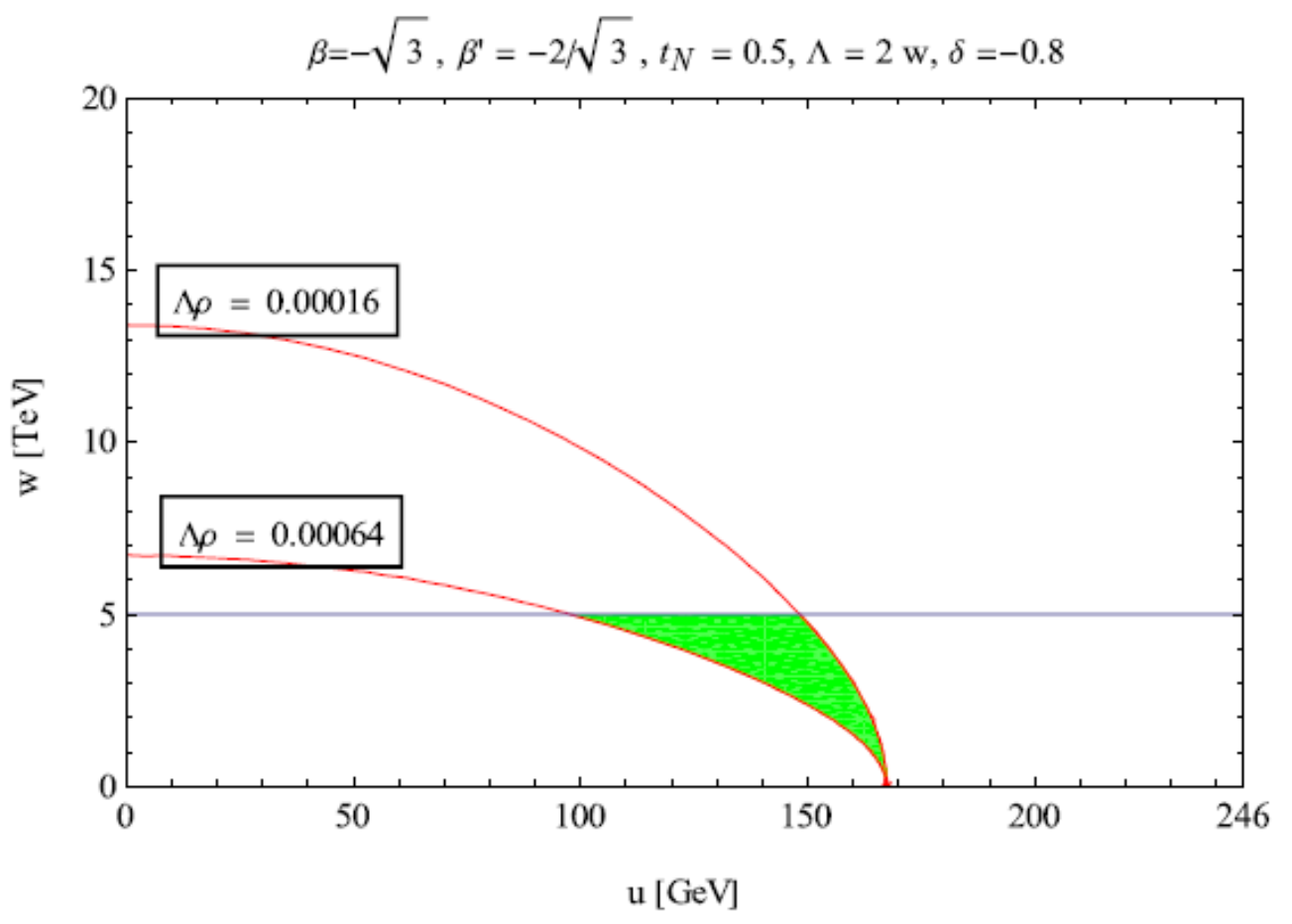}
\includegraphics[scale=0.4]{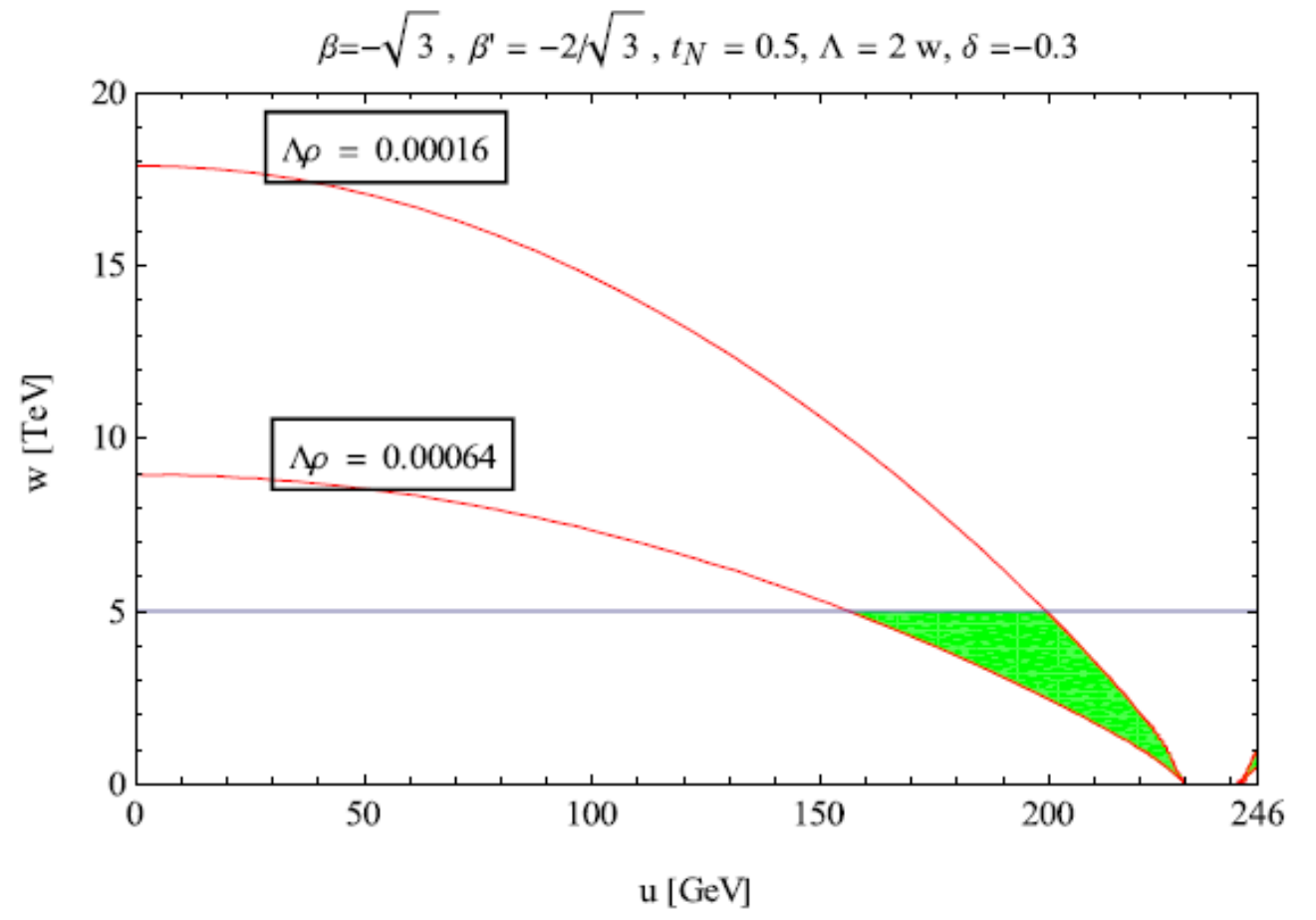}
\includegraphics[scale=0.45]{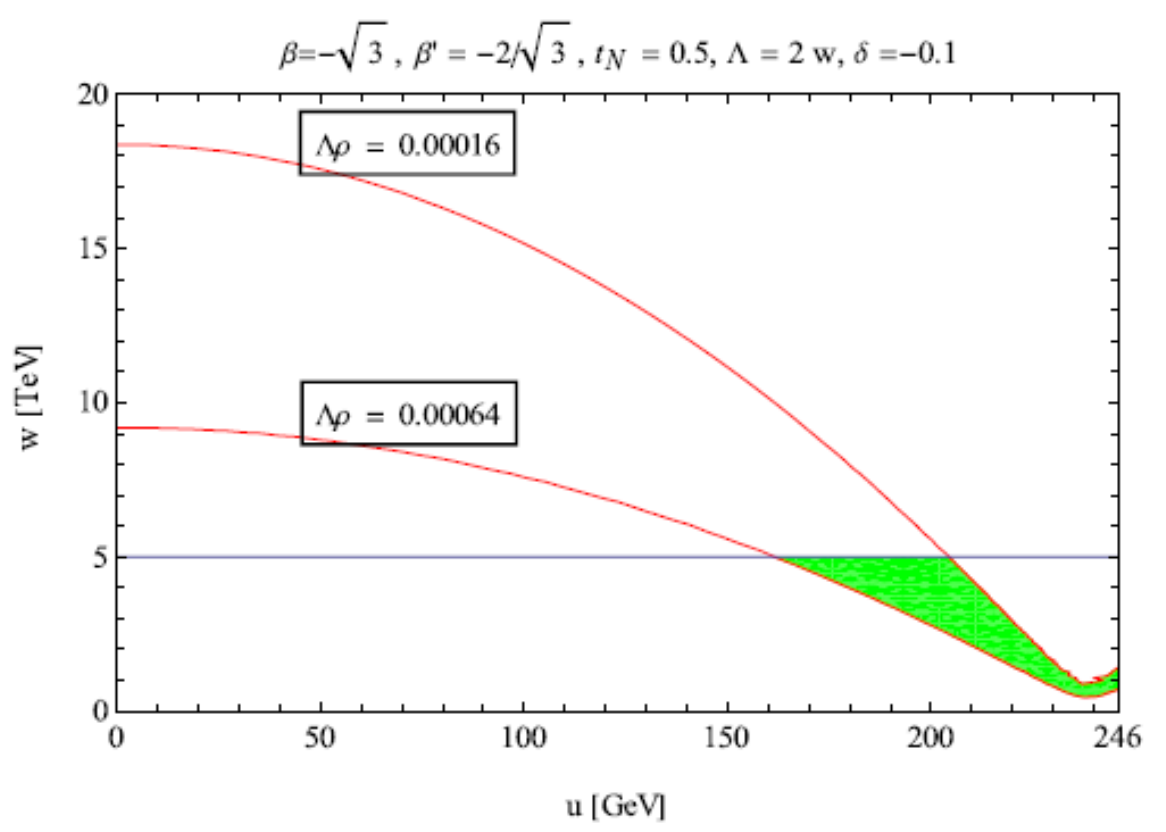}
\includegraphics[scale=0.4]{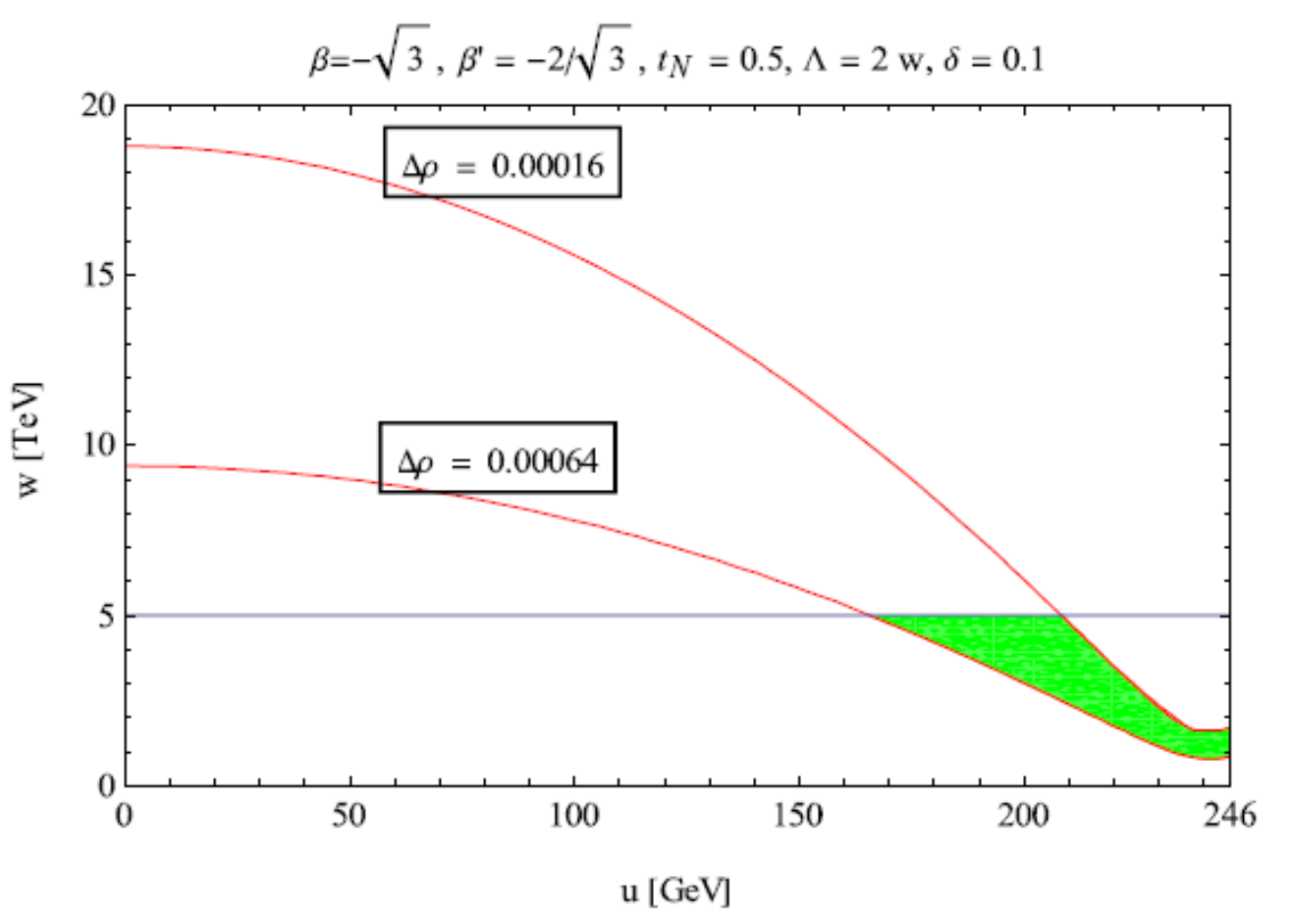}
\includegraphics[scale=0.4]{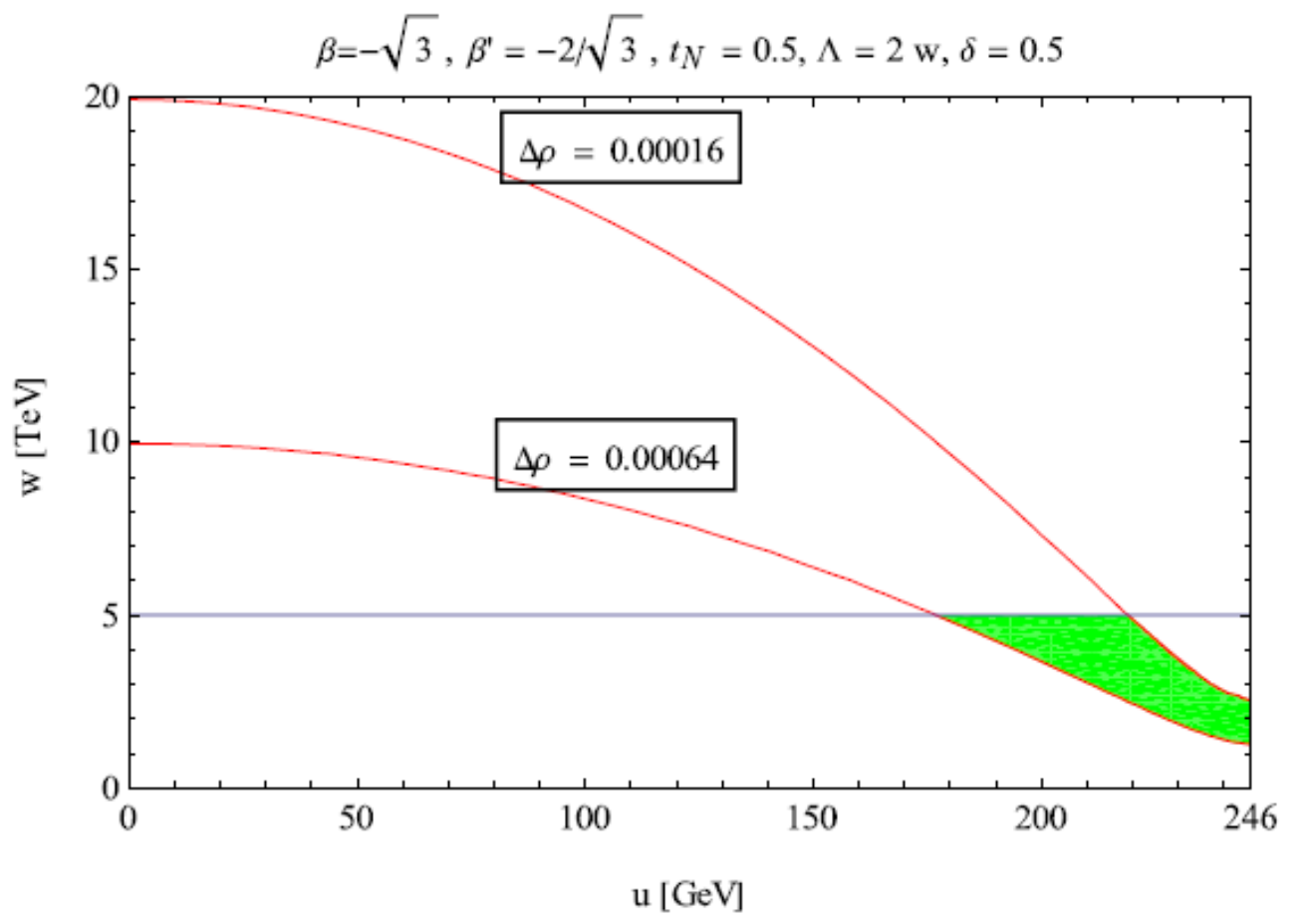}
\includegraphics[scale=0.45]{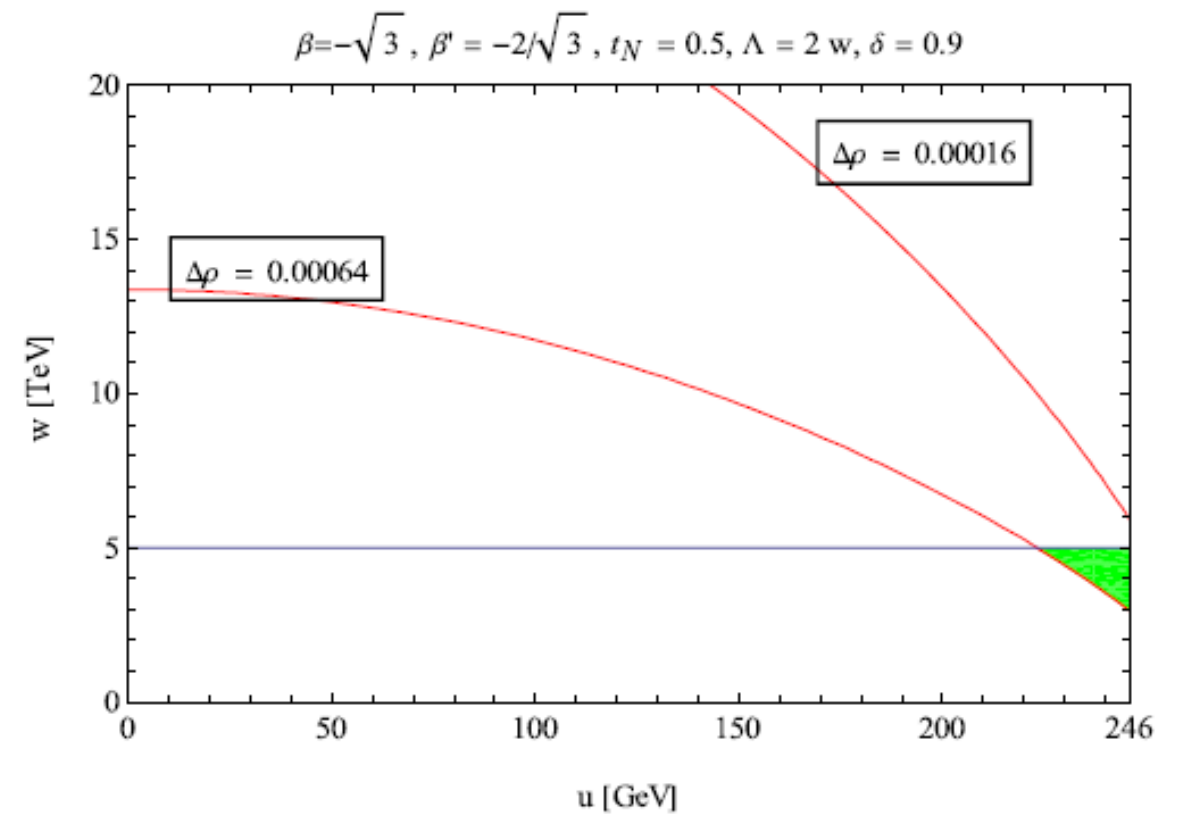}
\caption[]{\label{rho3311m} The $(u,w)$ regime that is bounded by the $\rho$ parameter for $\beta=-\sqrt{3}$, $\beta'=-2/\sqrt{3}$, $t_N=0.5$, and $\La=2w$. Here, the panels, arranging from left to right in raws and then from the top raw down to the bottom raw, are for $\delta=-0.8,\ -0.3,\ -0.1,\ 0.1,\ 0.5$, and $0.9$, respectively. In this case, the Landau pole, which is roundly $w=5$ TeV, is imposed.}
\end{center}
\end{figure}

In Fig. \ref{rho331}, we contour $\Delta \rho$ as a function of two variables $(u,w)$ for the case $\La\gg w$. It is clear from the figure that the bounds are independent of the new physics associated with $U(1)_N$ and the kinetic mixing term, which coincide with those of the 3-3-1 models. The cases of $\La\sim w$ are given in Figs. \ref{rho3311r}, \ref{rho3311s}, and \ref{rho3311m} for $\beta=-1/\sqrt{3}$, $\beta=1/\sqrt{3}$, and $\beta=-\sqrt{3}$, respectively. We see that the bounds on the new physics scales $w,\La$ increase when $\delta$ increase. Therefore, the kinetic mixing effect is important when the new physics is considered. It is noteworthy that in all cases of $\beta=-\sqrt{3}$ (also valid for the models of a seminal large $|\beta|$) the weak scales $u,v$ are very constrained since the new physics regime is limited below a low Landau pole (see also \cite{landau}). Furthermore, the new physics scales $w,\La$ from the mentioned figures are also subjected to other constraints, e.g. see the one in the next section, by which they would be larger than some TeV. These two extra bounds if applied will be neglected, which should be understood, for the following discussion to keep a simplicity.               
           
Further, the constraints from the $\rho$-parameter on the 3-3-1-1 breaking scales $w$, $\La$ also depend significantly on $u$ and can even approach zero for certain values of $u$, when $\delta$ is small and negative. Correspondingly, since $\Delta \rho$ is proportional to $\mathcal{E}$---the mixing of $Z$ with $Z'$ and $C'$---at the tree-level, the new physics is always decoupled from the standard model when $w,\La$ tend to the weak scales and then to zero, where we see that the mixing effects and $Z$-coupling corrections vanish. Apparently, this property is always protected at loop levels since this regime of the theory preserves a good custodial symmetry, $SU(2)_{L+R}$, as in the standard model. Therefore, we can close the 3-3-1-1 symmetry at the weak scales, which is similar to the 3-3-1 models studied in \cite{close3311}, as also seen from Fig. \ref{rho331}. The conclusion is valid for any $\beta,\ \beta'$, and $w$-$\La$ relation. Only if the kinetic mixing parameter is positive and large ($\delta\rightarrow 1$), such closing effects are relaxed and even lost. 

Generalizing the results in the second article of \cite{3311}, we obtain the standard model Higgs boson $H\simeq (u S_1+v S_2)/\sqrt{u^2+v^2}$, given at the leading order, $u,v\ll w,\La,-\mu$, where $S_1=\sqrt{2}\mathrm{Re}(\eta_1)-u$, $S_2=\sqrt{2}\mathrm{Re}(\rho_2)-v$, and $\mu$ is the triple coupling of $\eta$, $\rho$, $\chi$. The other scalars include nine massless Goldstone bosons as eaten by nine massive gauge bosons and ten new heavy Higgs bosons with masses in $w$, $\sqrt{|\mu w|}$, or $\La$ scale. The mass of $H$ can fit 125 GeV independent of $v/u$ ratio. At the leading order, the $H$ couplings coincide with those of the standard model, $\mathcal{L}\supset -\fr{m_f}{v_{\mathrm{w}}}\bar{f}f H+\fr{g^2v_{\mathrm{w}}}{2}(W^+_\mu W^{-\mu}+\fr{1}{2c^2_W}Z_\mu Z^\mu) (H+\fr{1}{2v_{\mathrm{w}}}H^2)$, where $m_f=-h_f\fr{u}{\sqrt{2}}$ for $f=t,d,s$, $m_f=-h_f \fr{v}{\sqrt{2}}$ for $f=b,e,\mu,\tau$, and $m_f=h_f\fr{v}{\sqrt{2}}$ for $f=u,c$. The modifications to those couplings due to the mixing of $H$ with new scalars are easily evaded since they are suppressed by $(u,v)/(w,\La,-\mu)$~\cite{3311}. We conclude that the standard model Higgs couplings and Higgs and fermion masses can be recovered at the leading order, without imposing any constraint on the ratio of the weak scales $v/u$, which is unlike the minimal supersymmetric standard model. The constraint on $v/u$ can only come from some of the following sources: (i) the $\rho$-parameter; (ii) when $\mu$ is low, by contrast; (iii) the perturbative limit of quark couplings with new scalars relevant to the top coupling, e.g. $\fr{m_t}{v_{\mathrm{w}}}\fr{v}{u}\bar{t}t H_1$ and $-\fr{m_t}{v_{\mathrm{w}}}\fr{v}{u}\bar{b}_L t_R H^-_2+H.c.$, where $H_1$ is orthogonal to $H$ while $H_2$ is a combination of $\rho_1,\eta_2$; and (iv) collider bounds on the new Yukawa, gauge, and scalar couplings. The last three cases are merely assumptions, which will be not considered in this work.            

The mixing of the standard model $Z$ boson with the new neutral gauge bosons will modify the well-measured couplings of the $Z$ with fermions. In the aforementioned basses, we have $Z=Z_1+\mathcal{E}_1 \mathcal{Z}'+\mathcal{E}_2 \mathcal{C}'$, $Z'=-\mathcal{E}_1 Z_1 + \mathcal{Z}'$, and $C'=-\mathcal{E}_2 Z_1+\mathcal{C}'$. Hence, the couplings of $Z_1$ to fermions include the corrections that come from the beginning $Z',\ C'$ couplings, which are proportional to $\mathcal{E}_1$ and $\mathcal{E}_2$, and obviously independent of the $\mathcal{Z'}$ and $\mathcal{C}'$ mixing angle. The various $\bar{f}fZ_1$ couplings have been examined \cite{pdg}, and the new physics contributions are safe if the mixing parameters are typically proportional to $10^{-3}$, by which we will take the bound $|\mathcal{E}_{1,2}| = 10^{-3}$ into account.        

\begin{figure}[!h]
\begin{center}
\includegraphics[scale=0.4]{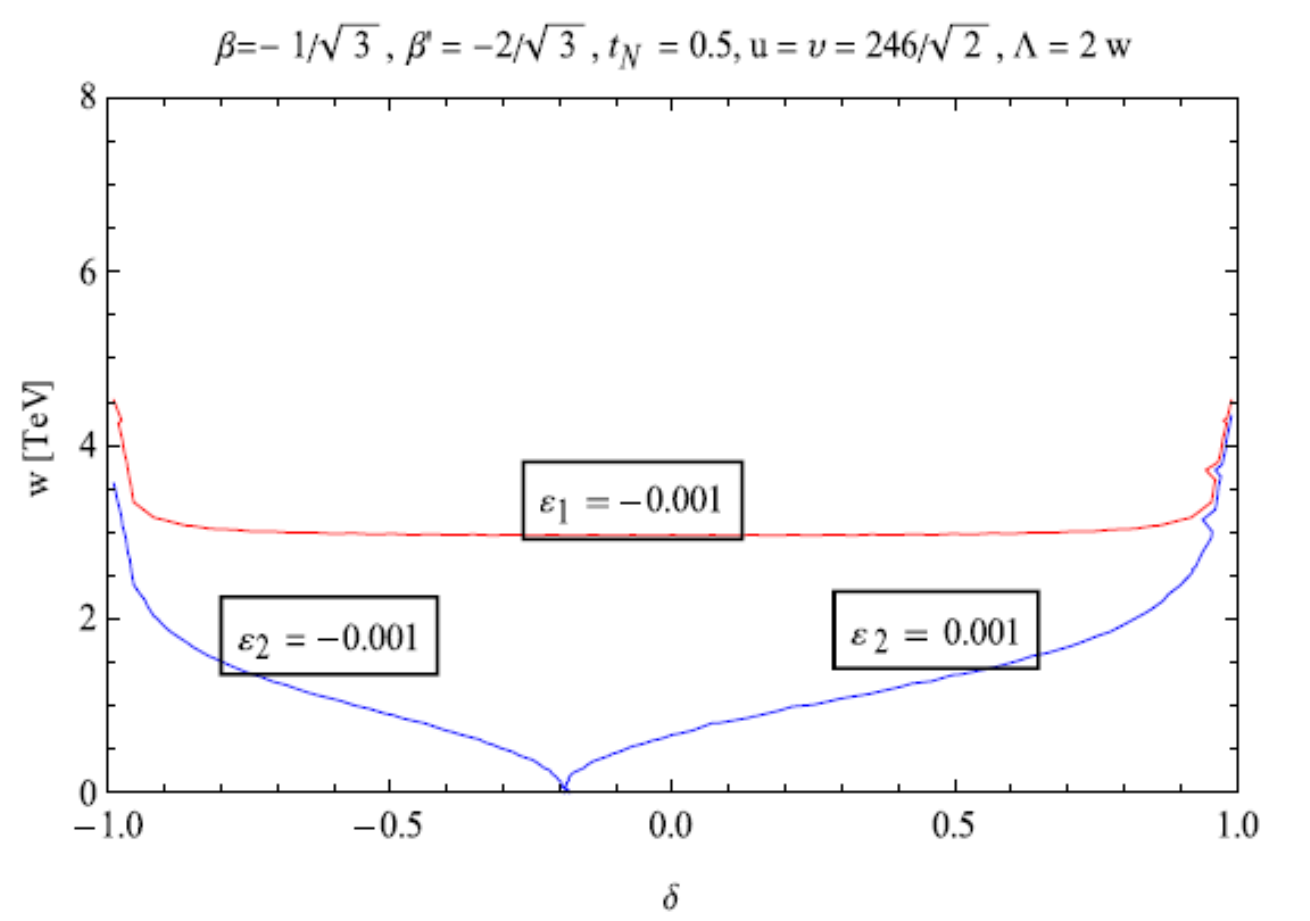}
\includegraphics[scale=0.45]{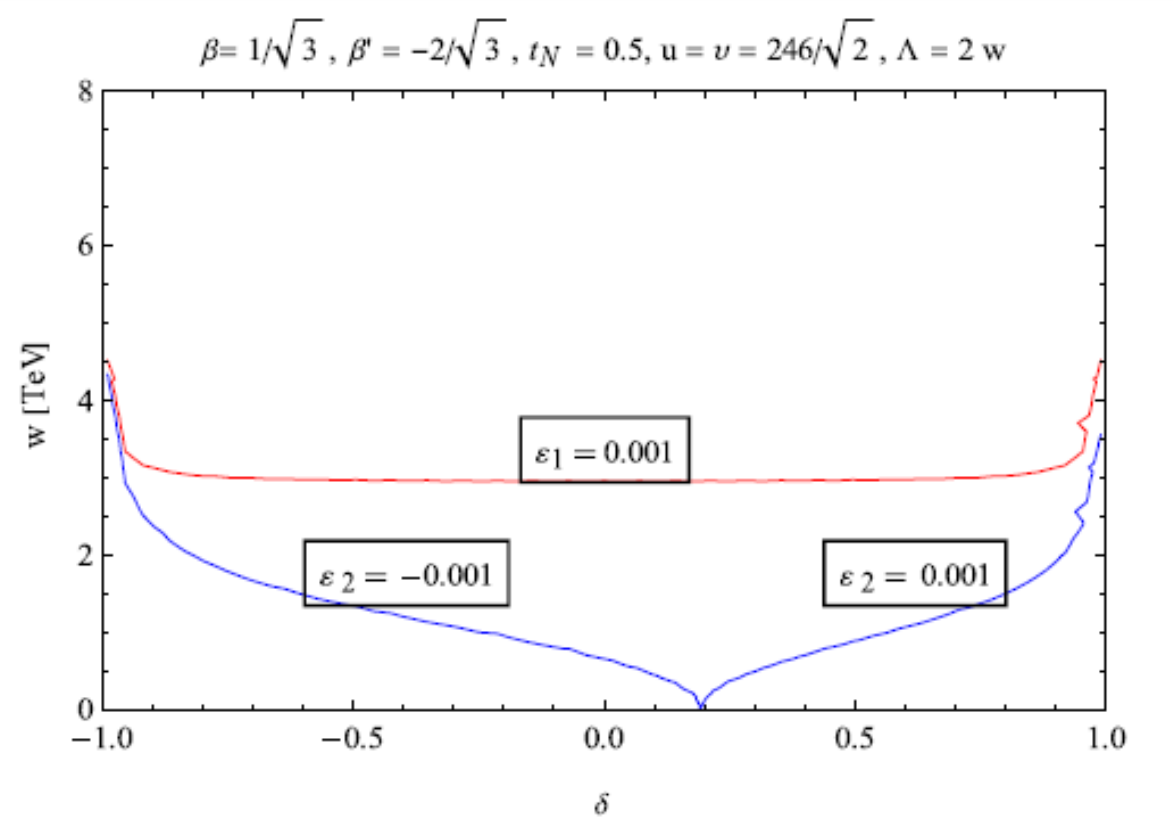}
\includegraphics[scale=0.4]{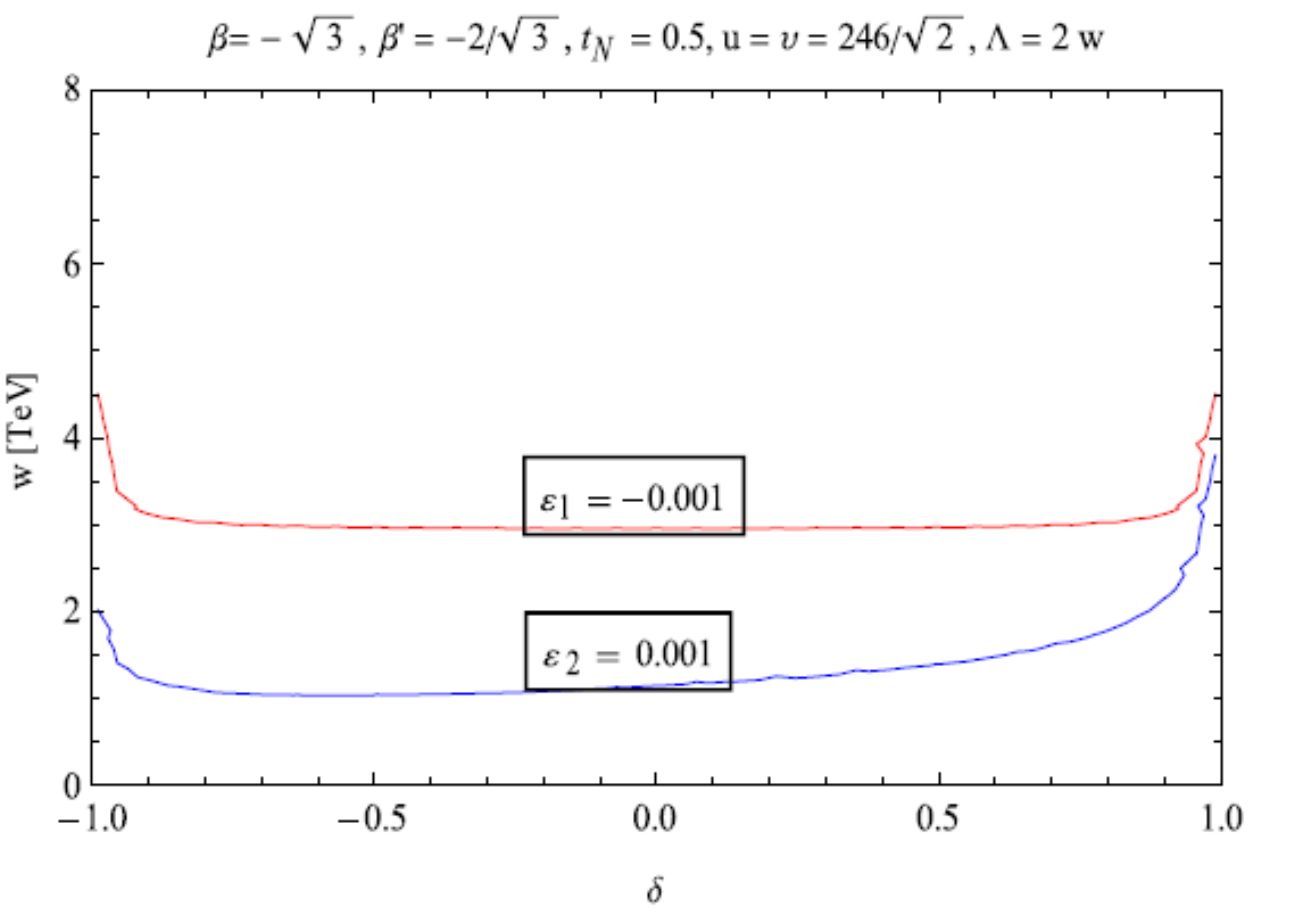}
\caption[]{\label{e12} The bounds on the new physics scales as functions of $\delta$, contoured by $|\mathcal{E}_{1,2}|=10^{-3}$, for the three kinds of the models $\beta=-1/\sqrt{3}$, $\beta=1/\sqrt{3}$, and $\beta=-\sqrt{3}$, respectively.}
\end{center}
\end{figure}

We observe that the sensitivity of the mixing parameters to the weak scales $u,v$ is only one term in $\mathcal{E}_{1}$ that is identical to the corresponding 3-3-1 model, since $u^2+v^2=(246\ \mathrm{GeV})^2$ is fixed. Also, if $\La\gg w$, $\mathcal{E}_2=0$, while $\mathcal{E}_{1}$ becomes that of the corresponding 3-3-1 model, $\mathcal{E}^0_1$. Therefore, these two cases are not investigated in this work, which should be well-understood. Our concern is the change of the new physics scales in terms of the kinetic mixing contribution, and for this case we can set $u=v$. The previous inputs, $\La=2w$, $t_N$, $\beta$, $\beta'$ are also used. In Fig. \ref{e12}, the viable new physics region is given above both lines of $\mathcal{E}_{1,2}$. Notably, the case of $\beta=-\sqrt{3}$ is subjected to the Landau pole limit, which implies that the new physics regime is very narrow.         

\section{Flavor-changing neutral currents}

The fermion generations are generally not repeated (or universal) under the $SU(3)_L\otimes U(1)_X\otimes U(1)_N$ symmetry, therefore there could be FCNCs. With the aid of  $X=Q-T_3-\beta T_8$ and $N=B-L-\beta' T_8$, the neutral currents take the form  
\be \mathcal{L}_{\mathrm{NC}}=-g\bar{F}\ga^\mu[T_3 A_{3\mu} + T_8 A_{8\mu}+t_X(Q-T_3-\beta T_8)B_\mu+t_N (B-L-\beta' T_8)C_\mu]F,\ee where $F$ is summed on all the fermion multiplets. It is clear that the leptons, including new particles $\nu_R$ and $k$, and the exotic quarks do not flavor change, because the corresponding flavors that potentially mix such as $(\nu_{1L},\nu_{2L},\nu_{3L})$, $(e_{1L},e_{2L},e_{3L})$, $(e_{1R},e_{2R},e_{3R})$, $(\nu_{1R},\nu_{2R},\nu_{3R})$, $(k_{1L},k_{2L},k_{3L})$, $(k_{1R}, k_{2R}, k_{3R})$, $(j_{1L},j_{2L})$, and $(j_{1R},j_{2R})$ are identical under the gauge charges, respectively. (Note that $j_3$ does not mix with $j_{1,2}$ due to the difference of electric charges). Additionally, the terms of $T_3$, $Q$, and $B-L$ are also not flavor changing, because all the repetitive flavor structures, including the mentioned ones and $(u_{1L},u_{2L},u_{3L})$, $(u_{1R},u_{2R},u_{3R})$, $(d_{1L},d_{2L},d_{3L})$, and $(d_{1R},d_{2R},d_{3R})$, are identical under those charges, respectively. Hence, the FCNCs are only associated with the ordinary quarks and $T_8$, by which we come with concerned interactions, 
\bea \mathcal{L}_{\mathrm{NC}}&\supset& -g\bar{q}_L\ga^\mu T_{8q}q_L (A_{8\mu}-\beta t_X B_\mu -\beta' t_N C_\mu), \eea
where $q$ denotes either up-type quarks $q=(u_1,u_2,u_3)$ or down-type quarks $q=(d_1,d_2,d_3)$, and $T_{8q}=\fr{1}{2\sqrt{3}}\mathrm{diag}(-1,-1,1)$ is the corresponding $T_8$ values. 

Let us work in the mass eigenstates, $q_{L,R}=V_{qL,qR}q'_{L,R}$, $q'=(u,c,t)$ or $q'=(d,s,b)$, and $(A_3\ A_8\ B\ C)^T=U(A,Z_1,Z_2,Z_3)^T$, which yields  
\bea \mathcal{L}_{\mathrm{NC}} &\supset& -\bar{q}'_L\ga^\mu (V^\dagger_{qL}T_{8q} V_{qL})q'_L(g_1Z_{1\mu}+g_2Z_{2\mu}+g_3 Z_{3\mu}),\crn
&\supset& -\fr{1}{\sqrt{3}}\bar{q}'_{iL}\ga^\mu q'_{jL} (V^*_{qL})_{3i}(V_{qL})_{3j}(g_1Z_{1\mu}+g_2Z_{2\mu}+g_3 Z_{3\mu}), \eea which implies the FCNCs for $i\neq j$. We also see that the photon always conserves flavors. The couplings of $Z_{1,2,3}$ are given by 
\bea && g_1=g\left[ -\fr{1}{\sqrt{1-\beta^2 t^2_W}}\mathcal{E}_1+\fr{1}{\sqrt{1-\delta^2}}(\beta' t_N-\delta \beta t_X)\mathcal{E}_2\right],\\
&& g_2=g\left[\fr{1}{\sqrt{1-\beta^2 t^2_W}}c_\xi +\fr{1}{\sqrt{1-\delta^2}}(\beta' t_N-\delta \beta t_X)s_\xi\right],\\ 
&& g_3=g_2(s_\xi\rightarrow -c_\xi,c_\xi\rightarrow s_\xi).\eea

The meson mixings are determined by the effective Lagrangian after integrating out $Z_{1,2,3}$, 
\be \mathcal{L}^{\mathrm{eff}}_{\mathrm{FCNC}}=\fr 1 3 (\bar{q}'_{iL}\ga^\mu q'_{jL})^2 [(V^*_{qL})_{3i}(V_{qL})_{3j}]^2\left(\fr{g^2_1}{m^2_{Z_1}} + \fr{g^2_2}{m^2_{Z_2}}+\fr{g^2_3}{m^2_{Z_3}}\right).\ee It is easily verified that the $Z_1$ contribution is negligible, because $(g^2_1/m^2_{Z_1})/(g^2_2/m^2_{Z_2}+g^2_3/m^2_{Z_3})$ is proportional to $(u^2,v^2)/(w^2,\La^2)$ that is suppressed at the leading order. Therefore, only $Z_2$ and $Z_3$ govern the FCNCs, which leads to 
\be \mathcal{L}^{\mathrm{eff}}_{\mathrm{FCNC}}\simeq \fr 1 3 (\bar{q}'_{iL}\ga^\mu q'_{jL})^2 [(V^*_{qL})_{3i}(V_{qL})_{3j}]^2\left(\fr{g^2_2}{m^2_{Z_2}}+\fr{g^2_3}{m^2_{Z_3}}\right).\ee    

The strongest bound comes from the $B^0_s$-$\bar{B}^0_s$ mixing, which is given by \cite{pdg}
\be \fr 1 3 [(V^*_{dL})_{32}(V_{dL})_{33}]^2\left(\fr{g^2_2}{m^2_{Z_2}}+\fr{g^2_3}{m^2_{Z_3}}\right)<\fr{1}{(100\ \mathrm{TeV})^2}.\ee
Supposing that the up-type quarks are flavor diagonal, we have the CKM factor $|(V^*_{dL})_{32}(V_{dL})_{33}|\simeq 3.9\times 10^{-2}$ \cite{pdg}. Hence, it follows  
\be \sqrt{\fr{g^2_2}{m^2_{Z_2}}+\fr{g^2_3}{m^2_{Z_3}}}<\fr{1}{2.25\ \mathrm{TeV}},\label{addfcncbb}\ee which yields $m_{Z_2}>2.25\times g_2\ \mathrm{TeV}$ and $m_{Z_3}>2.25\times g_3\ \mathrm{TeV}$, which are in the TeV order, assumed that the $g_{2,3}$ couplings are proportional to unity.       

\begin{figure}[!h]
\begin{center}
\includegraphics[scale=0.3]{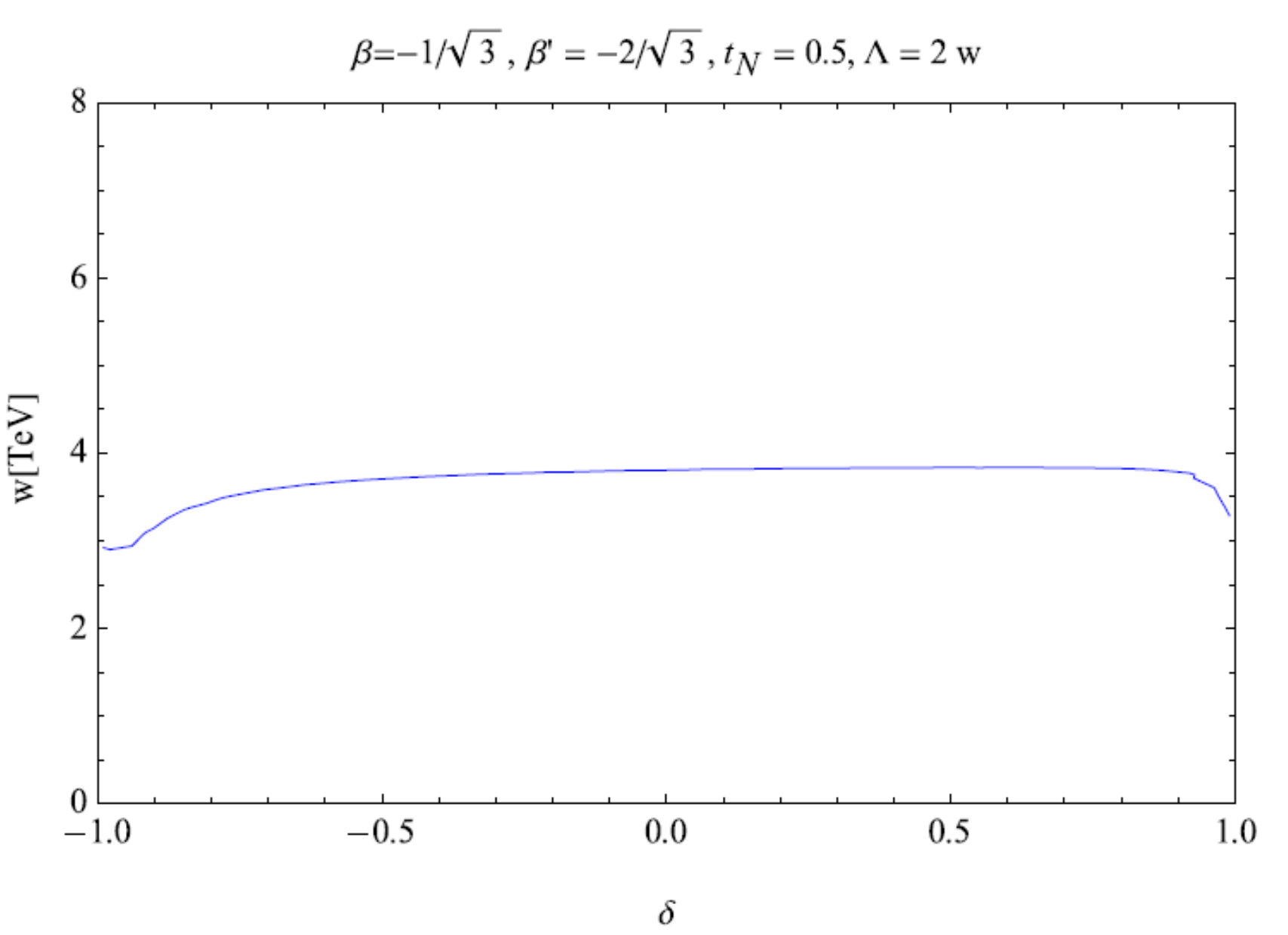}
\includegraphics[scale=0.4]{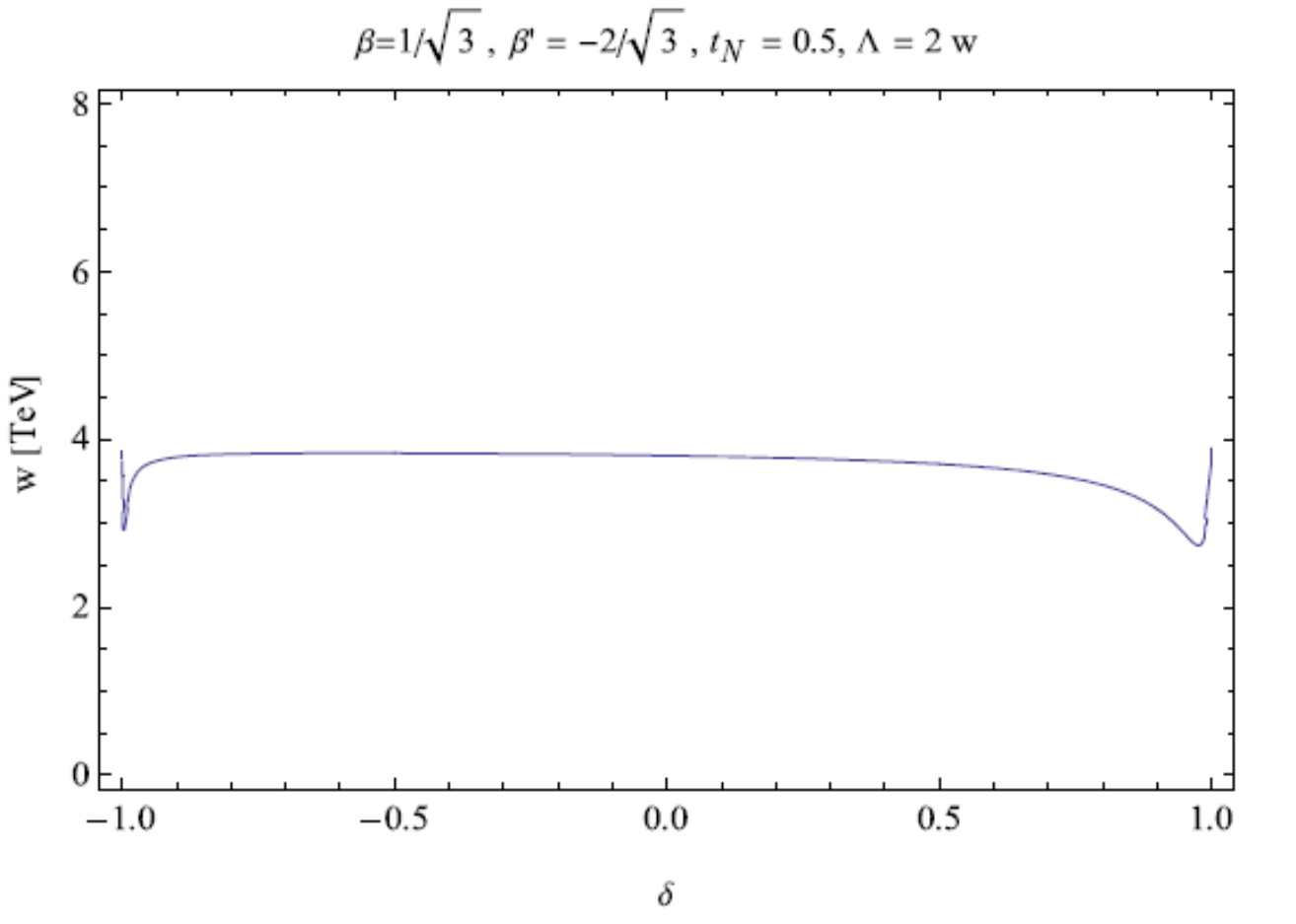}
\includegraphics[scale=0.4]{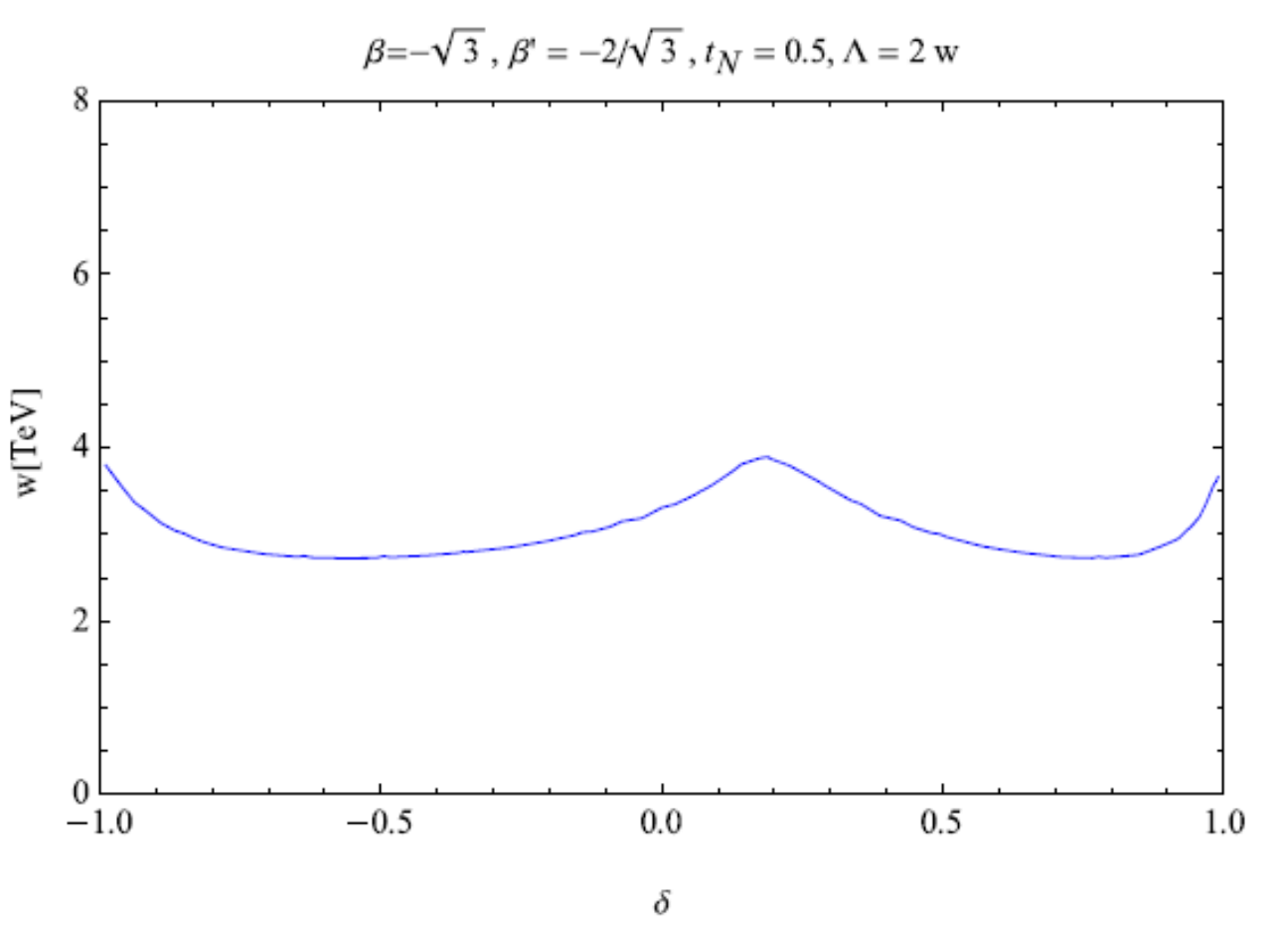}
\caption[]{\label{fcnc} The bounds on the new physics scales as functions of $\delta$, given by the constraint (\ref{addfcncbb}), for the three kinds of the models, $\beta=-1/\sqrt{3}$, $\beta=1/\sqrt{3}$, and $\beta=-\sqrt{3}$, respectively.} 
\end{center}
\end{figure}

Considering two conditions: 
\ben\item $Z_3$ is more superheavy than $Z_2$, i.e. $\La\gg w$. We have $m^2_{Z_2}\simeq \fr{g^2w^2}{3(1-\beta^2 t^2_W)}$, $m^2_{Z_3}\simeq 4g^2_N \La^2$, $\xi \simeq 0$, $g_2\simeq g/\sqrt{1-\beta^2 t^2_W}$, and $g_3\simeq -g(\beta' t_N-\delta \beta t _X)/\sqrt{1-\delta^2}$. The above bound becomes  $w>3.9\ \mathrm{TeV}$, which is given independent of $\beta$, $\beta'$, $g$, $g_X$, $g_N$ and $\delta$. This is also the common bound for every 3-3-1 model with arbitrary $\beta$. 
\item $Z_3$ and $Z_2$ are comparable in mass, so we take $\La =2 w$. The other inputs as given previously, $t_N$, $\beta$, and $\beta'$, are also used for this case. Since $u,v\ll w,\La$, the lhs of (\ref{addfcncbb}) depend only on the new physics scales, not on the weak scales. Additionally, the constraint (\ref{addfcncbb}) obeys the dependence of $w,\La$ bounds in terms of the kinetic mixing parameter, $\delta$, which is depicted in Fig. \ref{fcnc}. The figure shows that the new physics regime changes (although, slightly in a large region of $\delta$ for the left and middle panels), when $\delta$ varies. Those bounds are obviously lower than that given by the $\La\gg w$ case above.           
\een

\section{Conclusion}

Generally, if a theory contains two $U(1)$ gauge groups, the corresponding kinetic mixing term arises. Such term for the 3-3-1-1 model has been investigated, which is between the gauge bosons of $U(1)_X$ and $U(1)_N$, where these groups are used to define the electric charge $Q$ and baryon-minus-lepton charge $B-L$ as well as the necessary algebraic closure of these charges with $SU(3)_L$, respectively. The kinetic mixing modifies the neutral gauge boson spectrum of the original 3-3-1-1 model, which has been diagonalized in detail. The physical photon and $Z_1$ boson that are belong to the standard model have been identified. The new physical neutral gauge bosons $Z_2,Z_3$ that mainly relate to those of the 3-3-1 model and $U(1)_N$ have been achieved, which all are heavy in $w,\La$ scales, where $w$ is the 3-3-1 breaking scale, while $\La$ is the $B-L$ breaking scale. The mixing of $Z$ boson with new $Z',C'$ neutral gauge bosons are small, as governed by a seesaw like mechanism, in spite of the fact that the kinetic mixing parameter $\delta$ contributes and is finite. By contrast, the mixing between the new neutral gauge bosons $Z',\ C'$ is generally large, supplied by both sources, the 3-3-1-1 symmetry breaking and kinetic mixing term. In particular, the $Z'$-$C'$ mixing effect disappears when such contributions are canceled out, implying an interesting relation, $\delta=(\beta'g_N)/(\beta g_X)$, where $g_N$ and $g_X$ are the gauge couplings of $U(1)_N$ and $U(1)_X$, while $\beta'$ and $\beta$ are used to determine the $B-L$ and $Q$ embedding in the 3-3-1-1 symmetry, respectively.                 

The new physics regime is changed when the kinetic mixing contributes. The well-measured couplings of the standard model $Z$ boson are modified by the mixing parameters $\mathcal{E}_{1,2}$, which come from both the sources of the mixings, and they are properly suppressed by the mentioned seesaw like mechanism. The numerical investigation for the typical bounds, $\mathcal{E}_{1,2}=10^{-3}$, with a particular choice of the input parameters yields a solution for $w\equiv \La/2$ around 3 TeV, as given in the most range of $-1<\delta<1$. The $Z$-$Z'$ mixing ($\mathcal{E}_1$) is slightly varying in $\delta$, whereas the $Z$-$C'$ mixing ($\mathcal{E}_2$) is strongly sensitive to $\delta$. The new physics contribution to the $\rho$-parameter results from the tree-level mixings of $Z$ with $Z',\ C'$ bosons (due to both the sources, the 3-3-1-1 breaking and the kinetic mixing) as well as from the dominant one-loop corrections of the new non-Hermitian gauge bosons. Using the experimental data on $\rho$-parameter, the viable $(u,w)$ regions have been given, which are more sensitive to $\delta$ too. The new physics scale, $w$, is typically in TeV scale, while the weak scale, $u$, is correspondingly restricted. When $|\beta|$ is large, close to its bounds, the Landau pole might approach the weak scales, and in this case the VEVs $u,v$ are strongly definite, since their viable regimes are very narrow. Note that in this case, the new physics may be still decoupled from the standard model due to a good custodial symmetry $SU(2)_{L+R}$.                

We have calculated the flavor-changing neutral currents due to the discrimination of the third generation quarks from the first two as well as due to the mixing of the neutral gauge bosons. These currents are dominantly governed by the new $Z_{2,3}$ bosons, whereas the contribution of $Z_1$ boson is negligible. The experimental constraint on the $B^0_s$-$\bar{B}^0_s$ mixing sets the strongest bound on the new physics scales $(w,\La)$. If $\La\gg w$, the 3-3-1 breaking scale is bounded by $w>3.9$ TeV. If $\La =2 w$, the 3-3-1 breaking scale $w$ is around $3$--$3.5$ TeV. The new physics scales are obviously changed when $\delta$ varies. When $\beta$ is large so that the Landau pole is presented, close to TeV scale, the new physics regime due to both the constraints (the Landau pole and $B^0_s$-$\bar{B}^0_s$ mixing) may be very narrowed, as already seen for $\beta=-\sqrt{3}$.

Finally, we emphasize that the kinetic mixing effect must be taken into account when the new physics in the 3-3-1-1 model is examined. With the implication for dark matter, neutrino masses, cosmological inflation, and leptogenesis as well as the theoretical advantages over the known 3-3-1 models \cite{3311,d3311}, the current 3-3-1-1 model warrants further studies.             

\section*{Acknowledgments}

This research is funded by Vietnam National Foundation for Science and Technology Development (NAFOSTED) under grant number 103.01-2013.43.

\end{document}